\documentclass[fleqn,10pt,twocolumn]{wlscirep}
\usepackage[utf8]{inputenc}
\usepackage[T1]{fontenc}
\usepackage{wrapfig}
\usepackage{subcaption}
\usepackage{multirow}
\usepackage{multicol}
\usepackage{tikz}

\newcommand{\eg}{\emph{e.g.,~}}
\newcommand{\ie}{\emph{i.e.,~}}

\newcommand{\ket}[1]{\vert#1\rangle}
\newcommand{\bra}[1]{\langle #1\vert}

\newcommand{\dg}{^\dagger}
\newcommand{\nn}{\nonumber}
\newcommand{\pr}{{\rm Pr}}

\newcommand{\normf}[1]{\left\lVert #1 \right\rVert_F}
\newcommand{\normo}[1]{\left\lVert #1 \right\rVert}
\newcommand{\normt}[1]{\left\lVert #1 \right\rVert_1}
\newcommand{\normd}[1]{\left\lVert #1 \right\rVert_\diamond}
\newcommand{\E}[1]{\mathbb{E}\left\{ #1 \right\}}

\newcommand{\tr}{{\rm tr}}
\renewcommand{\k}{{(k)}}
\newcommand{\p}{{\bf\sf p}}

\long\def\comment#1{}

\title{Application Scale Quantum Circuit Compilation with Controlled Error}

\author[1,2,*]{Justin Kalloor}
\author[3]{Lucas Kovalsky}
\author[1,2]{Mathias Weiden}
\author[2]{John Kubiatowicz}
\author[1]{Ed Younis}
\author[1]{Costin Iancu}
\author[3,+]{Mohan Sarovar}
\affil[1]{Computational Research Division, Lawrence Berkeley National Laboratory, Berkeley, CA, 94702, USA}
\affil[2]{Department of Electrical Engineering and Computer Science, University of California, Berkeley, CA, 94702, USA}
\affil[3]{Quantum Algorithms and Applications Collaboratory, Sandia National Laboratories, Livermore, CA, 94550, USA}

\affil[*]{jkalloor3@berkeley.edu}
\affil[+]{mnsarov@sandia.gov}

\begin{abstract}
Compilation and optimization of quantum circuits are critical components in the execution of algorithms on quantum computers. These components must successfully balance two competing priorities: minimizing the number of expensive resources, such as two-qubit gates or arbitrary angle single-qubit rotations, and minimizing the approximation error of the compiled circuit to the ideal target unitary describing the quantum algorithm. We develop a practical workflow for managing and optimizing this tradeoff, which enables quantum circuit compilation and optimization at scales of hundreds of qubits. Our workflow is able to tackle circuits at such large scales while providing rigorous guarantees on circuit output error by leveraging circuit partitioning and the notion of averaging over circuit ensembles. We demonstrate our workflow on several benchmark algorithmic circuits acting on up to 380 qubits, and show that it can simultaneously achieve substantial reductions in resource-intensive gates and control output errors, offering a practical and scalable strategy for both near-term and fault-tolerant quantum computing.
\end{abstract}

\begin{document}

\flushbottom
\maketitle

\thispagestyle{empty}

\noindent Large-scale quantum computers have the potential to revolutionize
scientific and engineering computations. Critical to realizing
this potential is the compilation of a quantum
algorithm into a circuit composed of operations natively executable on a quantum computer, and subsequent optimization of this circuit taking into account constraints of the target hardware. These steps must be done for circuits operating on hundreds and thousands of qubits for applications of impact. We refer to this
as \emph{application scale} quantum circuit compilation and optimization. 

Compiled quantum circuits nearly always \emph{approximate} a target unitary prescribed by the quantum algorithm for multiple reasons. 
Firstly, utilizing approximate circuits
in order to reduce circuit depth, especially reductions in expensive
or noisy gates, is often desirable. In the near-term intermediate scale (NISQ) \cite{Preskill2018_NISQ} regime this is a
strategy to increase solution accuracy, while in the
fault-tolerant (FT) regime arbitrary angle rotations \emph{must be}
approximated via a fixed gate-set, and moreover, identifying and
removing insignificant but expensive gates (\eg very small angle
controlled rotations) is a strategy to increase circuit efficiency and
reduce execution time. Secondly, on the theoretical side, exact
circuit compilations are NP-hard to generate with increasing circuit
width \cite{botea2018complexity}, and thus approximations are likely
the best route to scalability.

While there exist many quantum circuit compilation tools \cite{qiskit,bqskit,tket,cirq,qdrift},
none are suitable for application scale compilation, which requires
efficient compilation of circuits with controllable approximations, acting on hundreds and thousands of
qubits. As quantum computing moves from the NISQ
era to early fault-tolerant \cite{PRXQuantum.5.020101} and ultimately full FT devices,
compilation and optimization must evolve to incorporate systematic
approximation mechanisms. Several challenges must be addressed: (i) {\it Quality of optimization:} circuit optimization must balance resource efficiency against approximation error, with tunable cost objectives; (ii) {\it Generality:} approximations must apply to circuits represented at multiple abstraction levels; NISQ-era, continuously parameterized gates and discrete FT gate sets such as Clifford+T; (iii) {\it Soundness;} approximation methods must rest on rigorously proven principles; (iv) {\it Rigorous guarantees:} optimized approximate compilations must come with guarantees on approximation error; and (v) {\it Scalability;} as mentioned above, circuit compilation and optimization tools must address application scale circuits. 

To meet these challenges, we develop and demonstrate an application scale quantum
circuit compilation and optimization workflow. Given an idealized or textbook circuit
expressed in either parameterized or discrete gates, our method
outputs an \emph{ensemble} of approximate compilations tailored to a
target device specification. The method supports arbitrary target error
levels and program execution consists of sampling circuits from the
ensemble and averaging measurement outcomes. This averaging defines a
quantum channel, and we derive rigorous error bounds for this channel
relative to the target unitary.

Our approach combines three main ingredients: (i) partitioning the
input circuit into non-overlapping subcircuits using binning
techniques that group adjacent gates into fixed-size blocks, (ii)
approximate compilation of each partition into an ensemble of
circuits, and (iii) optimization of the ensembles to guarantee circuit
output accuracy. The classical overhead of all steps scales
polynomially with circuit width and depth.

While these ingredients have been individually explored in previous work, their
integration into a complete, practical compilation workflow is our
main contribution. Circuit partitioning as a divide-and-conquer
strategy was previously proposed in Refs.~\cite{wu_qgo,Burt_2024}, but
these works neither analyzed approximation error nor established
output error guarantees.

The idea of ensemble averaging to reduce output error originates with
Campbell~\cite{PhysRevA.95.042306} and
Hastings~\cite{hastings2016turning}, who showed that properly designed
ensembles can suppress worst-case error quadratically
($\epsilon \rightarrow \epsilon^2$). Their methods, however, were
demonstrated only for single-qubit gates and relied on bespoke
compilation techniques (\eg Campbell's ensemble is
specified in terms of the Hamiltonians generating the gates or
circuits \cite{PhysRevA.95.042306}). Following this work, the insight
that averaging over several compilations of a circuit or gate can
reduce error has been applied in several contexts.  It has been
applied to generate randomized single qubit gate compilation
protocols \cite{Kliuchnikov_2023,Akibu_ACM_2024,yoshioka2024errorcrafting},
to define randomized versions of state preparation
circuits \cite{Akibue2024}, to randomize specific computational
primitives \cite{low2021halving}, and specific quantum algorithms,
such as product formula-based quantum
simulation \cite{campbell_qdrift,
Ouyang2020compilation,PRXQuantum.2.040305,huang_2023,PhysRevResearch.6.013224,PhysRevA.109.062431}
and quantum signal processing-based quantum
simulation \cite{martyn2024}.  Prior work by some of the authors
integrated circuit partitioning, synthesis, and
ensembles~\cite{patel_2021}, but lacked application-level scalability,
was primarily empirical, and did not derive worst-case error bounds or
methods for ensemble optimization.

In this work, we integrate theoretical results on error reduction
through circuit ensembles with practical compilation tools, producing
a scalable framework for compiling and optimizing quantum circuit ensembles with
rigorous worst-case output error guarantees.

\section*{Results}

Our workflow takes two inputs: a quantum circuit implementing a target unitary, $V$, and target output error $\epsilon^2$. This error could be with respect to the estimation of an observable or some distance metric on the output state of the circuit.  

The workflow builds on two principal strategies: (i) partitioning of the target circuit into smaller subcircuits, and (ii) averaging over ensembles of circuits to define an effective quantum channel. 
Partitioning decomposes the target circuit into a directed graph of
subcircuit blocks, each of which can be synthesized
independently. Ensemble averaging defines a quantum channel that
permits each subcircuit to be compiled only to accuracy $\epsilon$,
while rigorously ensuring that the overall circuit output error
remains bounded by $O(\epsilon^2)$. By combining circuit partitioning
with relaxed per-block synthesis accuracy, this approach substantially
improves compilation efficiency. Furthermore, it enables systematic
trade-offs in resource optimization, such as reducing the count of
two-qubit gates in NISQ implementations or minimizing the number of
non-Clifford gates in FT realizations. We view our workflow as the last step in a full compilation pipeline. Since we are generating approximate solutions, we can find further gate reductions on top of \emph{already optimized} circuits. 

\begin{figure}[t]
\begin{center}
    \includegraphics[width=\linewidth]{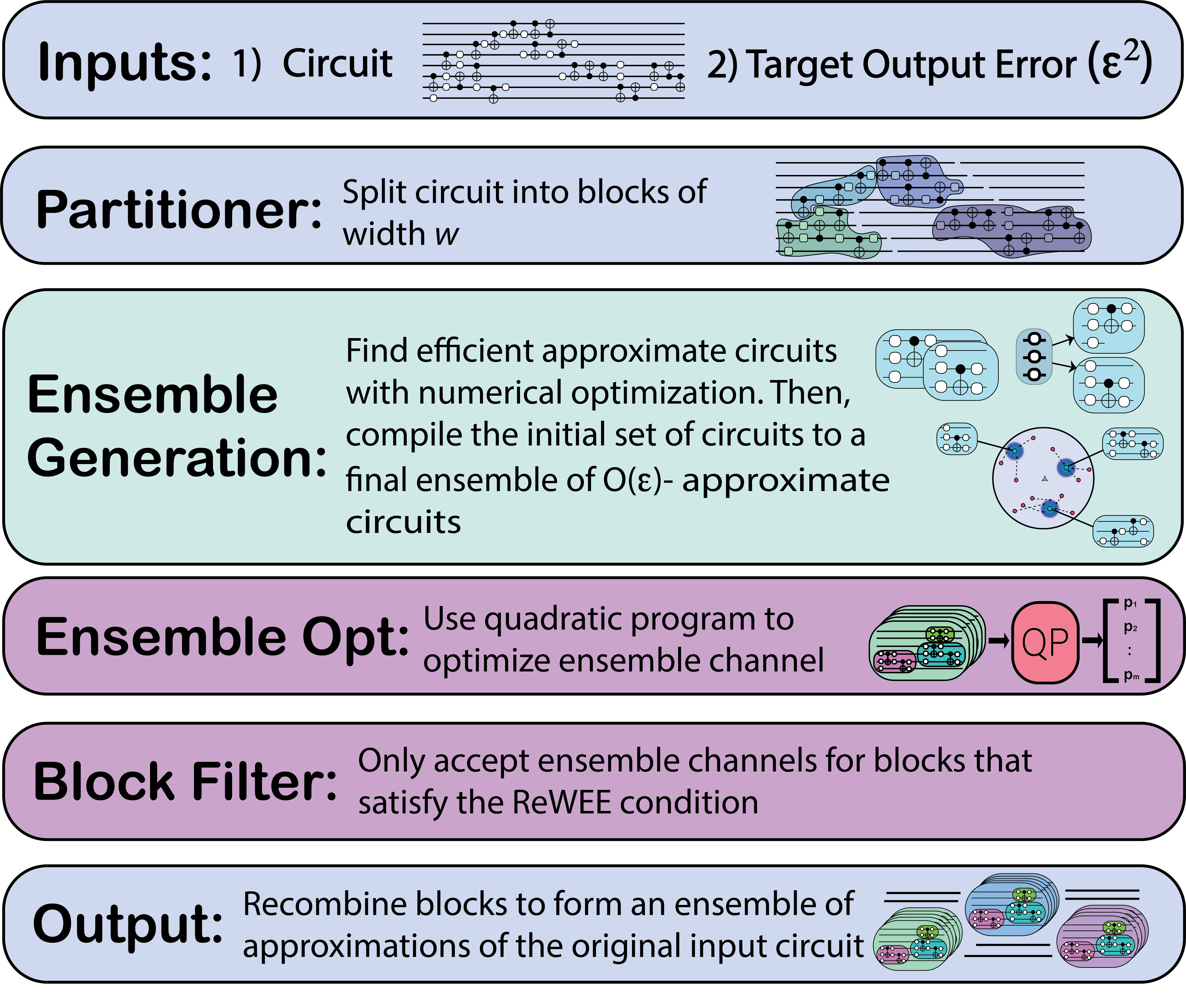}
\captionof{figure}{High level overview of our workflow. Each step is summarized in the text and discussed in detail in the Methods.}
\label{fig:resut_workflow}
\end{center}
\end{figure}

A high-level overview of the compilation workflow is presented in
Figure~\ref{fig:resut_workflow}, with detailed descriptions of each
step provided in the Methods section. We briefly describe each step here and the theoretical foundations that motivate it. 

The \emph{Partitioner} splits the input circuit into $K$ non-overlapping subcircuits, or \emph{blocks}. Then the \emph{Ensemble Generation} step creates an ensemble of circuits that are $\epsilon$-approximations to the target unitary for each block; \ie $||U^\k_i - V^\k||_F \leq \epsilon$, for $1\leq i \leq M^\k_{\rm synth}$, where $U^\k_i$ is the $i$th member of the ensemble for partition $k$, $V^\k$ is the target unitary for partition $k$, and $||A||_F \equiv \sqrt{\tr(A\dg A)}$ is the Frobenius norm. This norm is chosen at this step for its computational efficiency. Resource-aware
synthesis may be applied at this stage, \eg penalizing costly gates
such as arbitrary angle rotations. Additionally, this step aims increase the diversity of the ensemble for each block while maintaining the approximation error of each member of the ensemble below $\epsilon$. Next, the \emph{Ensemble Optimization} step solves a convex quadratic program to determine an optimal \emph{weighted} ensemble for each block; \ie weights $0 \leq p^\k_i \leq 1$ (with $\sum_{i=1}^{M^\k} p^\k_i = 1$) are computed that minimize the \emph{weighted ensemble error}: $\normf{\sum_{i=1}^{M^\k} p^\k_i U^\k_i - V^\k}$. 

\begin{figure}[t]
\centering
\includegraphics[width=0.8\linewidth]{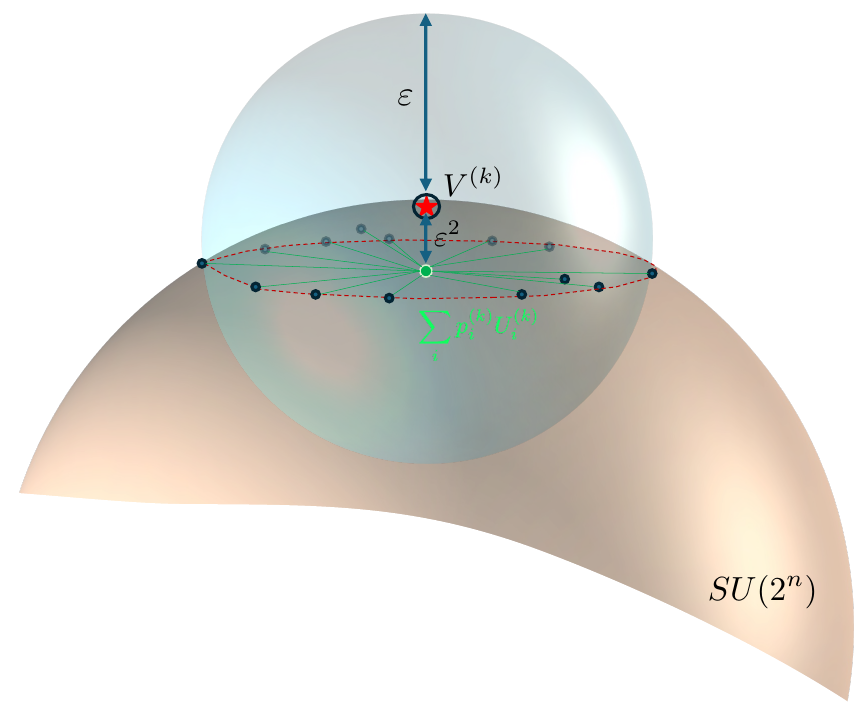}
\caption{
(a) Geometric intuition for the reduced Frobenius error condition. Ensemble members (blue dots) lie within an $\epsilon$-ball (light blue) around the target $V^{(k)}$ (red star). Their convex hull (dashed orange) contains the target, ensuring the weighted ensemble (green dot) lies within an $\epsilon^2$-ball (light green).
}
\label{fig:reduced_bias_combined}
\end{figure}

At the end of these steps, the critical property that we require is that the weighted ensemble error is quadratically reduced with respect to the original compilation tolerance $\epsilon$, \ie $\normf{\sum_{i=1}^{M^\k}p^\k_i U^\k_i - V^\k} \leq O(\epsilon^2)$. We refer to this as the \emph{reduced weighted ensemble error} (ReWEE) condition, and intuitively, it results from a synthesis algorithm that can generate a diversity of circuits whose convex hull contains a matrix that is closer to the target $V^{(k)}$ than any of the original circuits, as illustrated in Fig. \ref{fig:reduced_bias_combined}. A \emph{sufficient} condition for achieving ReWEE is an easily verifiable bias condition on the output of the \emph{Ensemble Generation} step: $\normf{\E{U_i^\k}-V^\k} \leq O(\epsilon^2)$, see Lemma 2 in Methods. The expectation above is taken with respect to the circuit distribution output by the Generation step.

Under the ReWEE condition, we show in the Methods that an immediate consequence of the \emph{Mixing Lemma} \cite{PhysRevA.95.042306,hastings2016turning}, is that the channel induced by averaging over the ensemble of circuits, $\mathcal{U}^\k[\rho] = \sum_{i=1}^{M^\k} p^\k_i U^\k_i \rho U_i^{(k)\dagger}$, has quadratically reduced error to the target channel, 
\begin{align}
    d_{\diamond}\left(\mathcal{U}^\k, \mathcal{V}^\k\right) \leq O(\epsilon^2),
    \label{eq:diamond_bound}
\end{align} 
where $\mathcal{V}^\k[\rho] = V^\k \rho V^{(k)\dagger}$ is the action of the target unitary, and $d_\diamond(\cdot,\cdot)$ is the diamond distance (defined in Methods), which has the interpretation as the worst-case discrepancy between the outputs of two channels \cite{watrous_2018}. 

The channel $\mathcal{U}^\k$ defines a discrete distribution over $M^\k$ circuits (set by $\{p^\k_i\}_{i=1}^{M^{(k)}}$ and denoted $\mathbb{P}_k$), and Eq. \eqref{eq:diamond_bound} says that averaging over circuits sampled from this distribution results in worst-case output error $O(\epsilon^2)$, compared to $\epsilon$ if any one of the members of ensemble is executed. This quadratic reduction in worst-case error is the key benefit to using the ensemble of compilations. 

Given optimized ensembles for all $K$ partitions, this defines a joint distribution $\mathbb{P} \equiv \mathbb{P}_1 \times ... \times \mathbb{P}_K$, from which circuits are sampled when executing the full-scale circuit. By the triangle inequality, the overall circuit error is bounded as $d_\diamond(\mathcal{U}^{(K)}\circ ... \circ \mathcal{U}^{(1)}, \mathcal{V}^{(K)}\circ ... \circ \mathcal{V}^{(1)}) \leq O(K \epsilon^2)$. The impact of the quadratic reduction in output error can be interpreted in two ways: (i) enabling coarser approximations at the subcircuit synthesis stage, or (ii) enabling application of the workflow to a larger number of partitions and thus larger circuits.

Note that if the \emph{Ensemble Generation} and \emph{Ensemble Optimization} steps are unable to find circuit ensembles that satisfy the ReWEE condition for some block, then the \emph{Block Filter} step replaces the ensemble channel for that block with the original circuit without approximations and sacrifices resource optimization in that block. Such filtering can occur if some blocks afford no resource optimization without incurring significant approximation error. Note that the output error bound is still guaranteed because the original circuit has zero approximation error.

\begin{table*}[t]
\centering
\scriptsize
\resizebox{\textwidth}{!}{%
\begin{tabular}{|l|r|r|r|r|r|r|}
\hline
Circuit & Reference & $\epsilon^2 = 10^{-2}$ & $\epsilon^2 = 10^{-4}$ & $\epsilon^2 = 10^{-6}$ & $\epsilon^2 = 10^{-8}$ & $\epsilon^2 = 10^{-10}$ \\
\hline
Heisenberg-7q & 360 & \textbf{259.1 (-28.0\%)} & \textbf{296.5 (-17.6\%)} & \textbf{323.9 (-10.0\%)} & 332.9 (-7.5\%) & \textbf{322.1 (-10.5\%)} \\
\hline
Fermi Hubbard-8q & 91 & 84.9 (-6.7\%) & 84.0 (-7.7\%) & 85.0 (-6.6\%) & 82.9 (-8.9\%) & 91 (0\%) \\
\hline
QAOA-10q & 60 & \textbf{53.5 (-10.8\%)} & \textbf{52.5 (-12.5\%)} & 54.4 (-9.3\%) & \textbf{51.6 (-14.0\%)} & 55.5 (-7.5\%) \\
\hline
Li-H Sim-12q & 4555 & \textbf{3729.4 (-18.1\%)}  & 4403.5 (-3.3\%) & 4405.9 (-3.3\%) & 4412.2 (-3.1\%) & 4450.4 (-2.3\%) \\
\hline
QFT Add-12q & 96 & 96 (0\%) & 96 (0\%) & 96 (0\%) & 96 (0\%) & 96 (0\%) \\
\hline
QAE-13q & 156 & 155.7 (-0.2\%) & 155.8 (-0.1\%) & 156 (0\%) & 156 (0\%) & 156 (0\%) \\
\hline
QPE-14q & 3444 & 3444 (0\%) & 3444 (0\%) & 3444 (0\%) & 3444 (0\%) & 3444 (0\%) \\
\hline
Mult-16q & 1108 & \textbf{963.4 (-13.0\%)} & \textbf{944.3 (-14.8\%)} & 1008 (-8.9\%) & 1021.5 (-7.8\%) & 1083.4 (-2.3\%) \\
\hline
Lattice Sim-17q & 124 & \textbf{107.4 (-13.4\%)} & \textbf{110.6 (-10.8\%)} & \textbf{107.4 (-13.4\%)} & \textbf{107.6 (-13.2\%)} & \textbf{112.9 (-9.0\%)} \\
\hline
QAE-33q & 1026 & \textbf{610.3 (-40.5\%)} & \textbf{724.9 (-29.3\%)} & \textbf{915.7 (-10.7\%)} & 1026 (0\%) & 1026 (0\%) \\
\hline
QAOA-148q & 1752 & 1751.0 (-0.1\%) & 1751.0 (-0.1\%) & 1751.0 (-0.1\%) & 1751.0 (-0.1\%) & 1752 (0\%) \\
\hline
Lattice Sim-380q & 3026 & \textbf{2934.6 (-3.0\%)} & 2944.8 (-2.7\%) & 2943.7 (-2.7\%) & 2941.6 (-2.8\%) & 2985.8 (-1.3\%) \\
\hline
\end{tabular}
}
\caption{CNOT Reductions across our set of benchmarks for a range of output error ($\epsilon^2$) values. For our reference circuit, we show results after performing a full peephole optimization in PyTket \cite{tket}, a state of the art exact compilation tool. We run our workflow on this optimized circuit. Significant CNOT reductions $\geq 10\%$ are marked in bold.}
\label{tab:nisq}
\end{table*}

\begin{table*}[t]
\setlength{\tabcolsep}{5pt}
\centering
\scriptsize
\begin{tabular}{|l|r | r|r | r|r | r|r | r|r | r|}
\hline
\multirow{2}{*}{Circuit} 
& \multicolumn{2}{c|}{$\epsilon^2 = 10^{-2}$} 
& \multicolumn{2}{c|}{$\epsilon^2 = 10^{-4}$} 
& \multicolumn{2}{c|}{$\epsilon^2 = 10^{-6}$} 
& \multicolumn{2}{c|}{$\epsilon^2 = 10^{-8}$} 
& \multicolumn{2}{c|}{$\epsilon^2 = 10^{-10}$} \\
\cline{2-11}
& \textbf{Ref.} & \textbf{Ens.} & \textbf{Ref.} & \textbf{Ens.} & \textbf{Ref.} & \textbf{Ens.} & \textbf{Ref.} & \textbf{Ens.} & \textbf{Ref.} & \textbf{Ens.} \\
\hline
Heisenberg-7q 
& 7240 & \textbf{1602.3 (-78\%)} 
& 11120 & \textbf{4061.3 (-63\%)} 
& 14800 & \textbf{6828.0 (-54\%)} 
& 18440 & \textbf{8800.7 (-52\%)} 
& 21960 & \textbf{10779.1 (-51\%)} \\
\hline
Fermi Hubbard-8q 
& 1358 & \textbf{807.0 (-41\%)} 
& 2068 & \textbf{1321.1 (-36\%)} 
& 2762 & \textbf{1804.5 (-35\%)} 
& 3492 & \textbf{2273.9 (-35\%)} 
& 4218 & \textbf{2749.3 (-35\%)} \\
\hline
QAOA-10q 
& 1180 & \textbf{565.3 (-52\%)} 
& 1840 & \textbf{954.2 (-48\%)} 
& 2386 & \textbf{1297.5 (-46\%)} 
& 3052 & \textbf{1683.2 (-45\%)} 
& 3650 & \textbf{2039.2 (-44\%)} \\
\hline
Li-H Sim-12q 
& 27570 & \textbf{20298.5 (-26\%)} 
& 44834 & \textbf{33119.9 (-26\%)} 
& 60076 & \textbf{44556.4 (-26\%)} 
& 76330 & \textbf{58768.0 (-23\%)} 
& 92204 & \textbf{71504.1 (-22\%)} \\
\hline
QFT Add-12q 
& 2879 & \textbf{2009.7 (-30.2\%)} 
& 4199 & \textbf{2737.3 (-34.8\%)} 
& 5725 & \textbf{3432.7 (-40.0\%)} 
& 6993 & \textbf{4612.7 (-34.0\%)} 
& 8543 & \textbf{6115.8 (-28.4\%)} \\
\hline
QAE-13q 
& 6339 & \textbf{3337.4 (-47\%)} 
& 9471 & \textbf{5166.4 (-45\%)} 
& 12637 & \textbf{7389.9 (-42\%)} 
& 15647 & \textbf{9699.6 (-38\%)} 
& 18953 & \textbf{13502.8 (-29\%)} \\
\hline
QPE-14q 
& 126100 & 121870.4 (-3\%) 
& 187450 & 179895.5 (-4\%) 
& 250954 & 238468.1 (-5\%) 
& 312372 & 294709.0 (-6\%) 
& 373568 & 351661.5 (-6\%) \\
\hline
Mult-16q 
& 43126 & 39583.3 (-8\%) 
& 63906 & \textbf{49069.6 (-23\%)} 
& 86372 & \textbf{60497.5 (-30\%)} 
& 106016 & 90447.7 (-15\%)
& 128880 & 119441.3 (-7\%) \\
\hline
Latt. Sim-17q 
& 3806 & \textbf{2514.6 (-34\%)} 
& 6086 & \textbf{4111.3 (-32\%)} 
& 7874 & \textbf{5433.0 (-31\%)} 
& 9906 & \textbf{6765.0 (-32\%)} 
& 12136 & \textbf{8434.1 (-31\%)} \\
\hline
QAE-33q 
& 31842 & \textbf{25444.7 (-20\%)} 
& 56840 & 46367.2 (-18\%) 
& 81828 & 70327.7 (-14\%) 
& 101446 & 87478.6 (-14\%) 
& 122258 & 103978.1 (-15\%) \\
\hline
QAOA-148q 
& 47440 & \textbf{34612.4 (-27\%)} 
& 72750 & \textbf{39842.4 (-45\%)} 
& 100146 & 81231.7 (-19\%) 
& 126206 & \textbf{90374.3 (-28\%)} 
& 152652 & \textbf{108960.3 (-29\%)} \\
\hline
Lattice Sim-380q 
& 85686 & 76452.8 (-11\%) 
& 138560 & 126252.6 (-9\%) 
& 179604 & \textbf{130030.8 (-28\%)} 
& 226730 & 194379.3 (-14\%) 
& 277804 & 242570.8 (-13\%) \\
\hline
\end{tabular}
\caption{T gate reductions across our set of benchmarks for a range of output error ($\epsilon^2$) values. The reference circuit is generated with a PyTket optimized circuit for which all $R_z$ gates are approximately compiled into a fixed gateset by Gridsynth to error $\epsilon^2$. Significant T gate reductions $\geq 20\%$ are marked in bold.}
\label{tab:ft}
\end{table*}

\subsection*{Demonstration of resource reduction with controlled output error}
\label{sec:demos}
In the following, we demonstrate gate count reduction while controlling output error for a diverse set of benchmarks, including key quantum subroutines (Adders, Mults, QAEs, and Modular Exponentiation), Hamiltonian simulations (LiH, Fermi-Hubbard, Heisenberg, and Lattice Gauge Simulation), and optimization algorithms (QAOA). The circuit widths for these benchmarks vary from 7 qubits to 380 qubits. See Supplementary Information (SI), Section 1, for more details on each of the benchmarks. The implementations of the components in the workflow, Fig. \ref{fig:resut_workflow}, are discussed in Methods. 
For NISQ architectures, we compile to CNOT and arbitrary single-qubit gates, minimizing CNOT count. For FT architectures, we compile to single-qubit Clifford, CNOT, and $R_z$ gates, and then use Gridsynth \cite{ross_sellinger} to approximate the $R_z$ rotations as a sequence of $H, S$ and $T$ gates. We compare performance against \emph{optimized} reference circuits that are generated using PyTKet \cite{tket} with full peephole optimization. For the FT case, PyTKet compiles and optimizes into Clifford and $R_z$ gates and we again use Gridsynth to approximate the $R_z$ rotations. Note that while in the NISQ case, we compare against a reference circuit that exactly implements the target unitary, in the FT setting the reference is also an approximate circuit. In order to have fair comparison, in our workflow we use Gridsynth to synthesize each $R_z$ gate in a circuit with $n_Z$ $R_z$ gates to error $\frac{\epsilon}{n_Z}$, whereas in the reference each $R_z$ gate is synthesized to error $\frac{\epsilon^2}{n_Z}$. This ensures that the overall circuit/channel error is $O(\epsilon^2)$ in both cases. Unless otherwise stated, all numerical demonstrations use partitions consisting of subcircuits acting on four qubits; the rationale for this choice is discussed in the Methods section.

Tables \ref{tab:nisq} and \ref{tab:ft} summarize the gate count reductions achieved by our workflow for the benchmark suite.  
Table \ref{tab:nisq} reports CNOT counts for the exact (reference) compilation and the average CNOT counts for the circuit ensembles generated by our workflow for various output error bounds $\epsilon^2$.  
In most cases, we achieve reductions over the exact compilation, in some instances significantly (marked in bold). There are, however, notable exceptions. In the NISQ case, we see no improvement for the Multiplier, Fermi Hubbard, QPE, and QFT Adder circuits. In these cases, the exactly optimized circuits produce circuits with an extremely "rigid" structure. Across all blocks, our synthesis techniques were unable to find significant CNOT reductions while maintaining low approximation errors. As we reduced our error budget, our synthesizer would only find one or two reduced circuits. While these circuits did have fewer CNOTs, they were not sufficient to quadratically reduce the weighted ensemble error. We provide examples of such blocks in the SI, Section 2. 

Table \ref{tab:ft} presents results for the FT workflow.  
Here, we report $T$-gate counts for both the reference and our workflow, for various target output error values. Our approach reduces $T$-gate counts for all benchmarks and all target error values. This is primarily because we can leverage a higher approximation error for $R_z$ synthesis than the reference compilation.

Having demonstrated resource reductions, we now show that our approach also controls the output error of the compiled circuits. 
Direct numerical calculation of the diamond distance between the channel output by our approach and the ideal circuit is infeasible for our benchmarks. Instead, we use the fact that the diamond distance upper bounds both observable errors and output state errors, and we evaluate these more tractable quantities.

\begin{figure*}[ht]
    \centering
    \begin{subfigure}{0.49\textwidth}
        \centering
        \includegraphics[width=\linewidth]{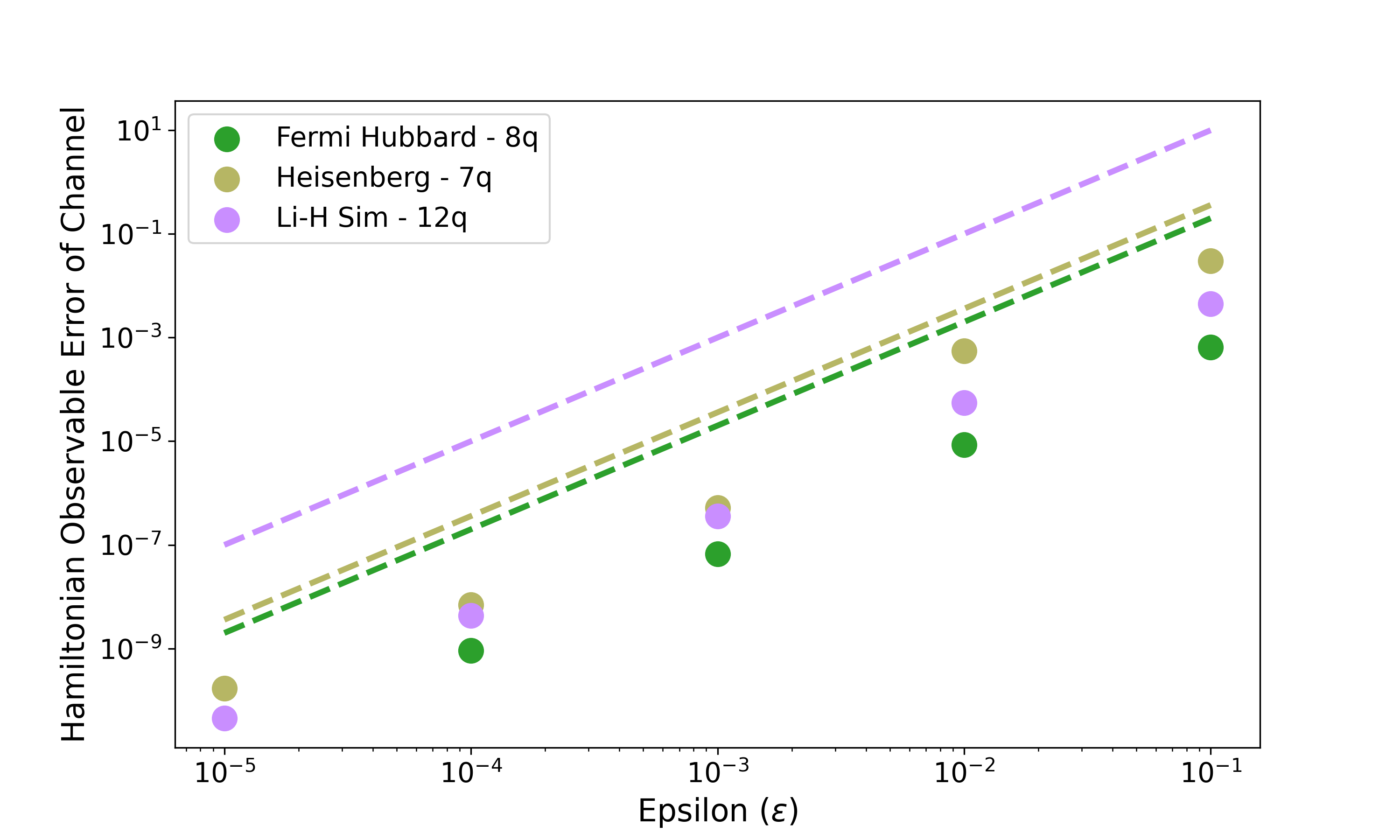}
    \end{subfigure}
    \begin{subfigure}{0.49\textwidth}
        \centering
        \includegraphics[width=\linewidth]{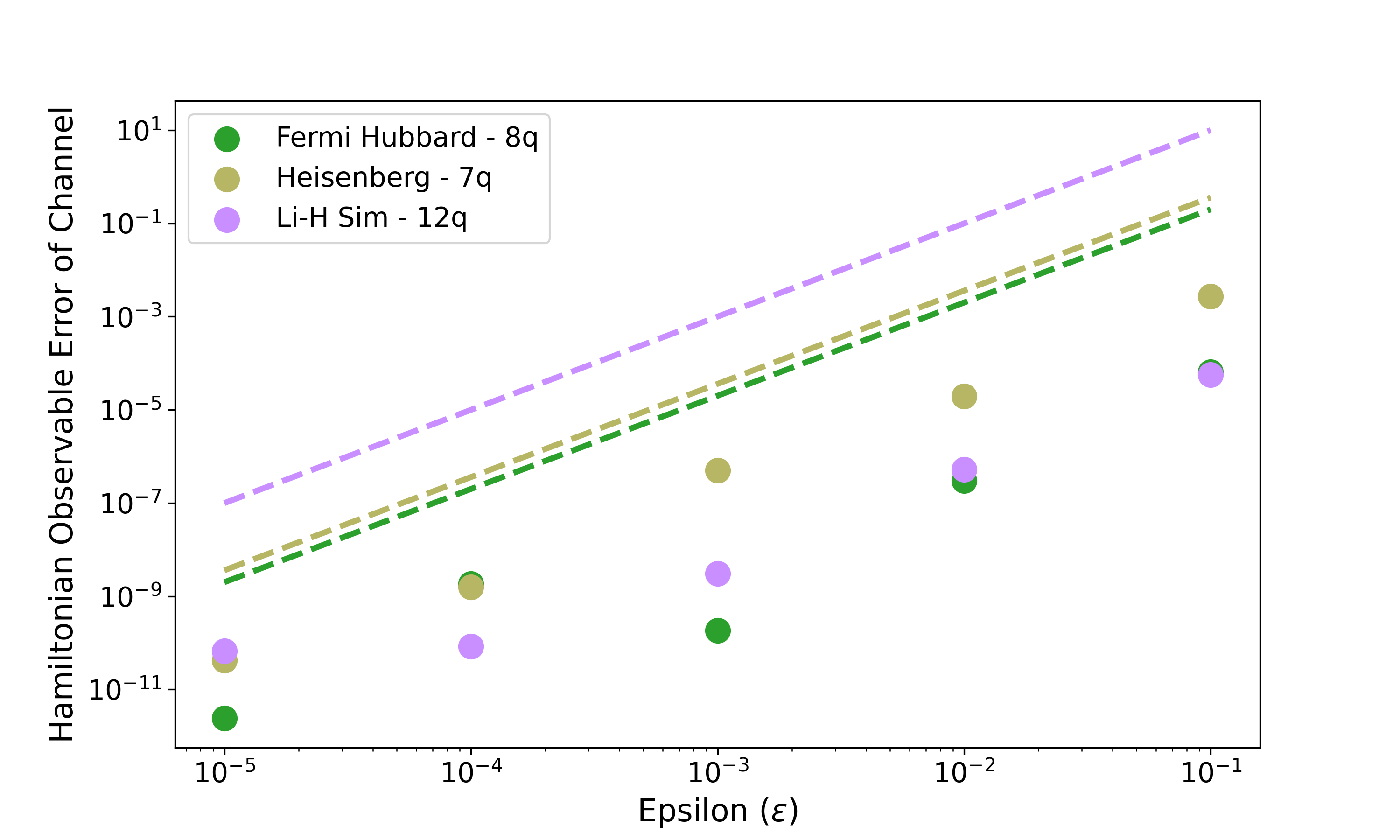}
    \end{subfigure} \\
    \caption{Observable errors (y-axis) for Fermi–Hubbard Model, LiH and Heisenberg benchmarks, as a function of the compilation error tolerance $\epsilon$.  
    The left panel shows NISQ compilation results; the right panel shows FT compilation results. For each benchmark, we draw a corresponding dotted line at $K\epsilon^2$ where $K$ is the number of blocks in the benchmark. Across all benchmarks, the observable error is bounded by this line.}
    \label{fig:obs_scaling}
\end{figure*}

Figure \ref{fig:obs_scaling} shows the observable error for three benchmark applications where observable measurement is appropriate.  
For each benchmark, we compute the expected value of an observable using the ideal circuit unitary, denoted $O_{\mathrm{ideal}}$ (the specific observables for each benchmark are described in the SI, Section 1).  
We then compute the same observable using the channel induced by the ensemble of compiled circuits produced by our workflow, for a range of target errors,  $O_{\mathrm{ens}} = \sum_{p_i \in \mathbb{P}} p_i O(U_i)$,
where $p_i$ are the probabilities assigned to circuits $U_i$ in the ensemble, the blocks of each $U_i$ are synthesized to Frobenius norm error $\epsilon$, and $O(U_i)\equiv\bra{\psi_0}U_i\dg O U_i \ket{\psi_0}$.  
To isolate the effects of compilation error, we neglect shot noise in both $O_{\mathrm{ideal}}$ and $O_{\mathrm{ens}}$.  
We expect the observable error $|O_{\mathrm{ideal}} - O_{\mathrm{ens}}|$ to scale as $O(K\epsilon^2)$, where $K$ is the number of blocks in the partitioning of the input circuit.  
Indeed, Fig. \ref{fig:obs_scaling} confirms this scaling for both NISQ and FT compilation workflows.

\begin{figure*}[ht]
    \centering
    \begin{subfigure}{0.48\textwidth}
        \centering
        \includegraphics[width=\linewidth]{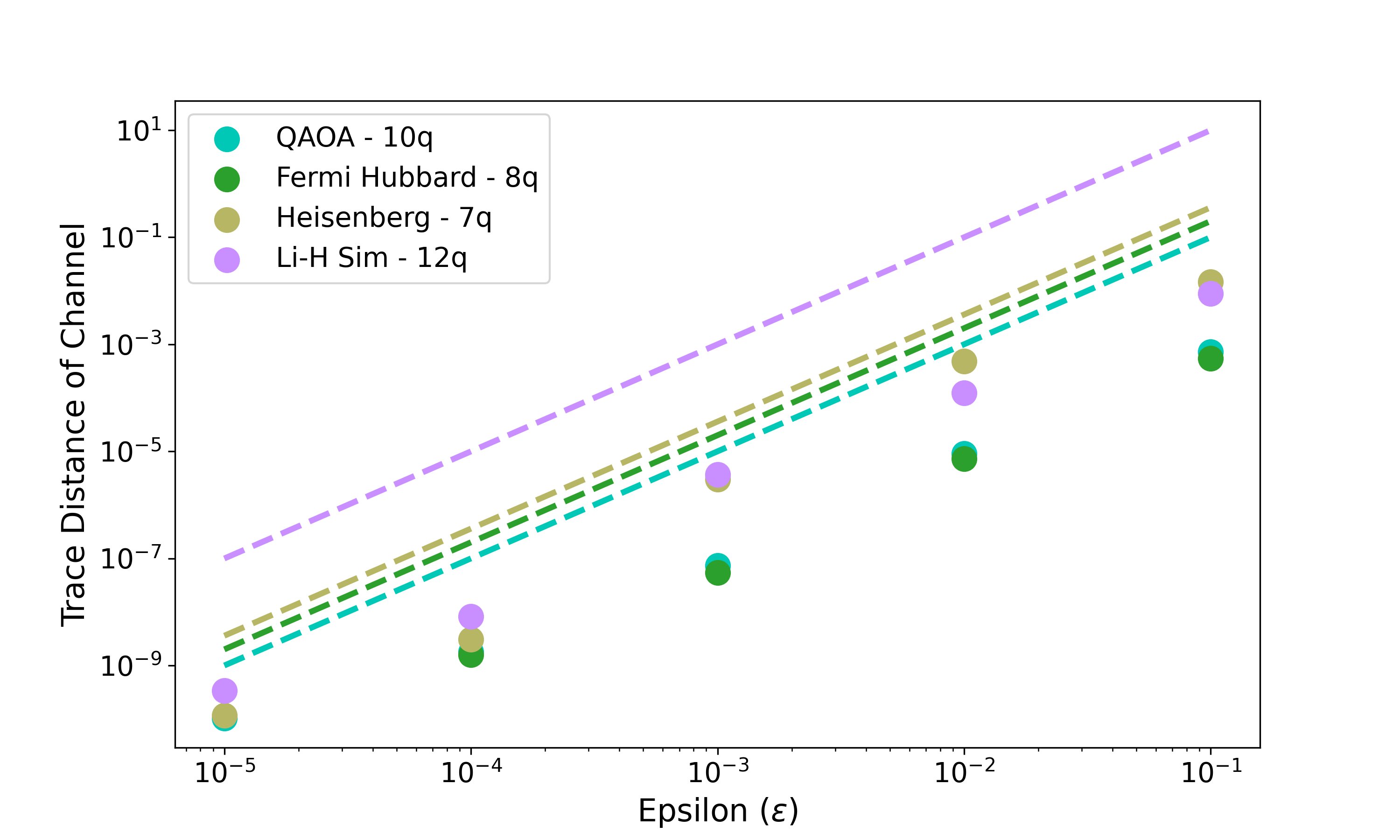}
    \end{subfigure}
    \begin{subfigure}{0.48\textwidth}
        \centering
        \includegraphics[width=\linewidth]{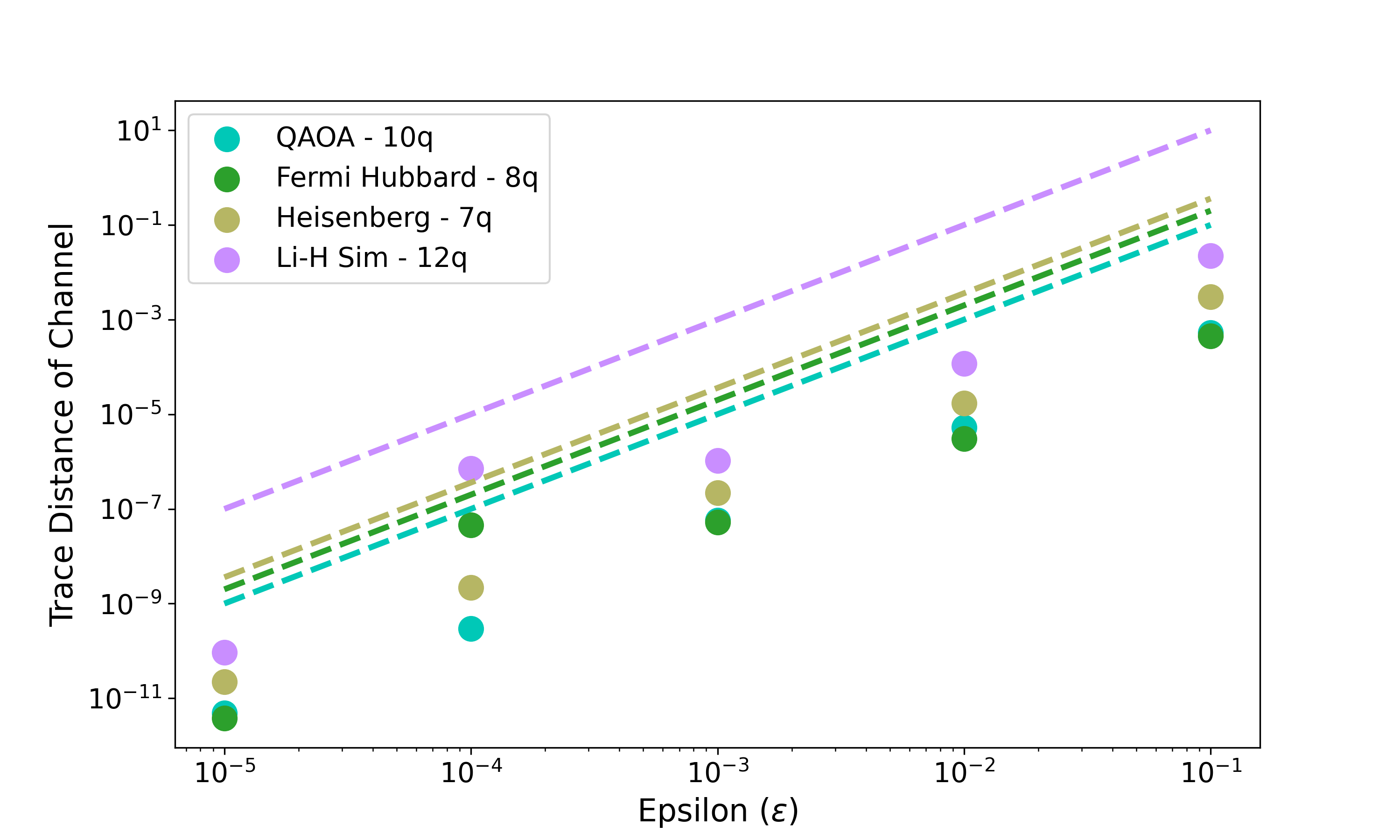}
    \end{subfigure} \\
    \caption{Output state errors (y-axis) for the QAOA, Fermi Hubbard, Heisenberg, and LiH benchmarks. For each benchmark and each $\epsilon$, we generate a set of 10 random input density matrices and choose the density matrix that maximizes the output trace distance with respect to the original circuit. This metric serves as a proxy for the worst-case distance (diamond distance) which is intractable to solve. We plot this maximum trace distance (y-axis) between ensemble and ideal outputs as a function of $\epsilon$ (x-axis). The left panel shows NISQ results; the right shows FT results. For each benchmark, we draw a corresponding dotted line at $K\epsilon^2$ where $K$ is the number of blocks in the benchmark.
    In all cases, the trace distance is bounded by this line.}
    \label{fig:trace_distance_scaling}
\end{figure*}

An alternative error metric that is tractable to compute is the trace distance between the output state of the ensemble and the ideal output state for a given input state, $\frac{1}{2}\left\| \rho_{\mathrm{ens}} - \ket{\psi_{\mathrm{ideal}}}\bra{\psi_{\mathrm{ideal}}} \right\|_1$,
which upper bounds the error of any observable for that fixed input state.  
For each benchmark, we prepare a set of 10 random input density matrices and compute the corresponding output density matrices after applying the ensemble channel. We take the maximum trace distance over these random inputs as a proxy for the worst-case channel error, which cannot be computed exactly. Figure \ref{fig:trace_distance_scaling} shows that this proxy also scales as $O(K\epsilon^2)$ for both NISQ and FT workflows.

For a comprehensive picture of the error scaling our framework achieves, see the SI, Section 3, where we present statistics of the weighted ensemble error and ensemble bias across all blocks for all benchmarks.

\begin{figure}[ht!]
    \centering
    \includegraphics[width=1\linewidth]{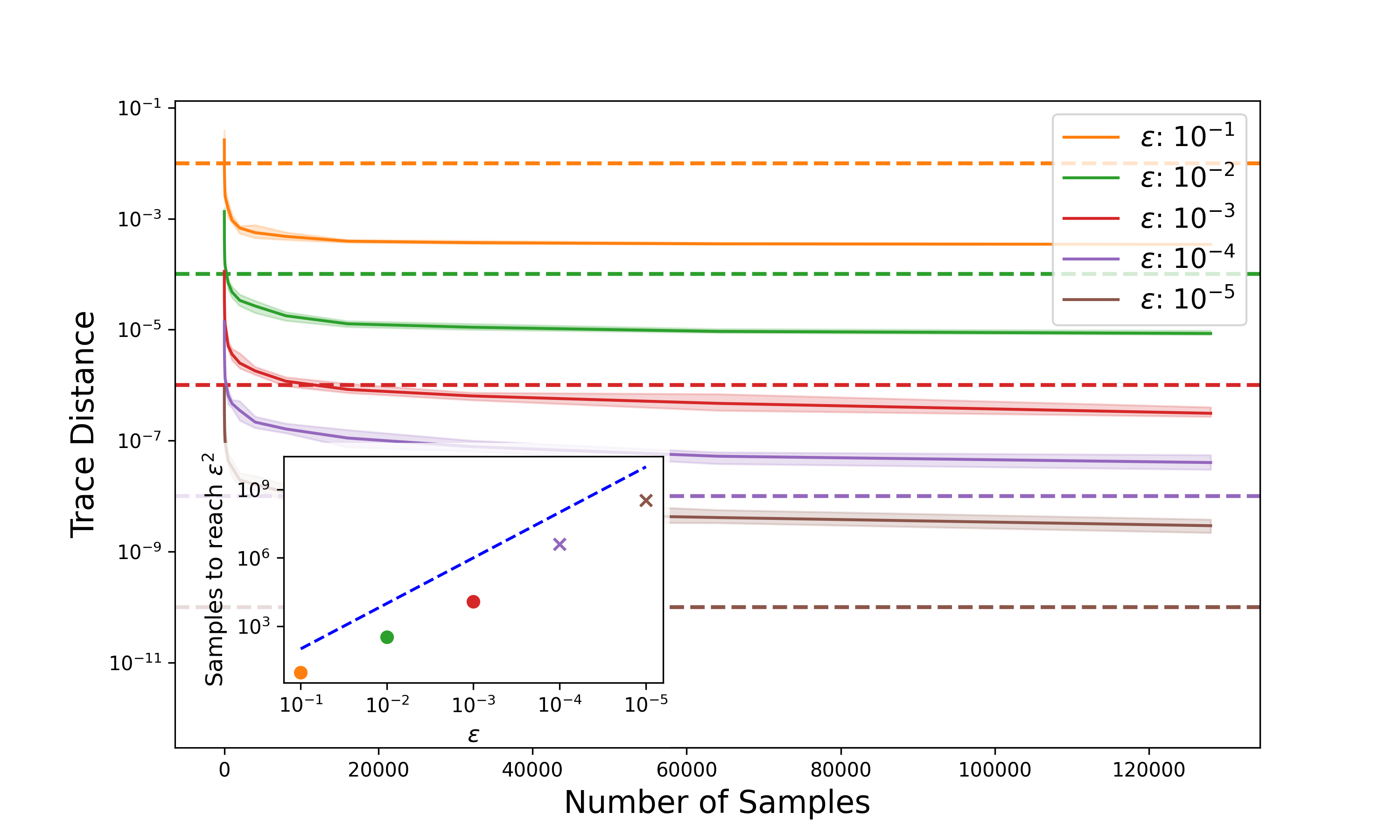}
    \caption{Convergence of output state error, as measured by trace distance, for the 7-qubit Heisenberg benchmark as a function of the number of samples drawn from the ensemble channel.  Dotted horizontal lines indicate $\epsilon^2$ for each target $\epsilon$. We run 10 different trials at each sample and plot the average (solid line) as well as the full range (area). The inset figure plots the number of samples required until the sampled empirical channel, $\hat{\mathcal{U}}[\rho]$, achieves an average output state error $\epsilon^2$ from the true channel, $\mathcal{V}[\rho]$. The number of samples required for $\epsilon \in \{ 10^{-4}, 10^{-5} \}$ are extrapolated values. The dotted blue line is equal to $\epsilon^{-2}$, and empirically upper bounds the number of samples required.
    }
    \label{fig:td_convergences}
\end{figure}

\subsection*{Sample convergence of approximate ensembles}

An important practical consideration is the \emph{sample complexity}—the number of circuits that must be executed to effectively implement the unital channel specified by a circuit ensemble. Lemma 3 in Methods proves a lower bound on the number of samples required to implement an ensemble channel to error $O(\epsilon^2)$. Specifically, given an ensemble channel $\mathcal{U}[\rho] = \sum_{i=1}^M p_i U_i \rho U_i\dg$ that approximates a target unitary channel $\mathcal{V}[\rho] = V \rho V^{\dagger}$ to error $O(\epsilon^2)$, \ie $d_\diamond(\mathcal{V}, \mathcal{U})\leq O(\epsilon^2)$, we show that the empirical channel $\hat{\mathcal{U}}[\rho] = \frac{1}{T}\sum_{i=1}^T U_{\sigma(i)} \rho U_{\sigma(i)}\dg$, with $\sigma(i)\in \{1,...,M\}$, defined by sampling and averaging over $U_i$ according to the distribution $\{p_i\}$, converges in the sense $\pr\left\{ d_\diamond(\mathcal{V}, \hat{\mathcal{U}}) < O(\epsilon^2) \right\} > 1-\delta$, given
\begin{align}
    T \geq \frac{2v + \frac{2}{3}R \left(\epsilon^2/(2d^2)\right) }{\left(\epsilon^2/(2d^2)\right)^2}\ln \frac{2d^2}{\delta},
\end{align}
where $d$ is the dimension of $V, U_i$, and $v$ and $R$ are measures of variance and maximum deviation of members of the ensemble (defined in Methods). This lower bound formally scales as $O(1/\epsilon^4)$, however in practice, the scaling with $\epsilon$ is gentler since $v$, the measure of variance in the ensemble, scales as $O(\epsilon^2)$.

We study the rate of convergence of empirical channels for one of our benchmarks in Figure \ref{fig:td_convergences}. Again, since diamond distance is intractable to compute, we compute the trace distance of output states (maximized over a sample of ten random input states) for the 7-qubit Heisenberg circuit compiled to the NISQ gate set as the number of samples in the empirical channel, $T$, increases. For each $\epsilon$, a dotted line marks the corresponding $\epsilon^2$ target.  
As shown in Fig. \ref{fig:trace_distance_scaling}, the optimized unital channel achieves this scaling for this benchmark; here, we observe how quickly it is approached as the sample size grows. Fitting and extrapolating from this data, the rate of convergence for output trace distance in this case, is $O(1/\epsilon^c)$, with $c<2$. In the SI, Section 4 we present statistics of the variance and maximum deviation measures, $v$ and $R$, for this benchmark and others, that provides evidence that $v$ scales as $\epsilon^2$ in practice.

\section*{Discussion}

We have developed a framework for compilation and optimization of application scale quantum circuits that can negotiate the trade off between reducing resource requirements and providing rigorous guarantees on output error. We demonstrated this framework across a diverse set of benchmark quantum algorithm circuits, using specific tools (\eg BQSKit, GridSynth) for components of our framework, and showed consistent reductions in resource-intensive gates -- CNOTs for NISQ architectures and $T$ gates for FT architectures -- while simultaneously controlling output errors.  
For benchmarks where the weighted ensemble error is quadratically reduced (\ie scales as $O(\epsilon^2)$, when circuits composing the ensemble are synthesized to approximation error $\epsilon$) for a significant fraction of circuit blocks, we observe resource reductions exceeding $40\%$ in some cases, without exceeding the prescribed error tolerance.  
Output state error and output observable errors scale as $O(\epsilon^2)$, in agreement with theoretical predictions, confirming that the quadratic error suppression extends from the block level to the full compiled circuit.  
These results demonstrate that ensemble-based approximate compilation can simultaneously deliver meaningful resource savings and provable error guarantees, making it a viable strategy for both near-term and fault-tolerant quantum computing regimes.

Our demonstrations used specific, extant tools to implement components of our workflow such as circuit partitioning and circuit synthesis. Our numerics validate the performance of these implementations. However, we emphasize that the overall framework, summarized in Fig. \ref{fig:resut_workflow}, and its theoretical foundations are the central contribution of this work. This framework allows for future development and maturation of components, especially the circuit synthesis and diversification components, that might allow for even better negotiation of the trade-off between resource reduction and control of output errors. 

\section*{Methods}

\subsection*{Theory}
The theoretical foundations of our compiler workflow are built on two results that we present in this section. First, we introduce some notation. We use the following matrix norms to talk about approximation errors:
\begin{enumerate}
	\item $\normf{A} = \sqrt{\tr(A\dg A})$, the Frobenius norm. 
	\item $\normo{A} = \sigma_{\rm max}(A)$, the operator or spectral norm. 
	\item $\normt{A} = \tr(\sqrt{A\dg A}) = \sum_{i=1}^m \sigma_i(A)$, the Schatten 1-norm or trace norm. Here, $A$ is assumed to be a $m\times m$ square matrix.
\end{enumerate}
In addition, we will use the diamond norm for quantum channels,
\begin{align}
	\normd{\mathcal{E}} = \sup_X \{ \normt{(\mathcal{E}\otimes \mathcal{I})(X)} ~;~ \normt{X}\leq 1 \},
\end{align}
where $\mathcal{I}$ is an identity channel of the same dimension as $\mathcal{E}$. This norm induces a \emph{diamond distance} between channels, $d_\diamond(\mathcal{E}, \mathcal{E}') = \frac{1}{2}||\mathcal{E} - \mathcal{E}'||_{\diamond}$.

Let $V$ be a target unitary on $n$ qubits, and $U_i$ be a set of approximate compilations of $V$ provided by a compiler. The compilation workflow can have probabilistic elements, and therefore we think of $U_i$ as independent, identically distributed (\emph{i.i.d.}) samples from some distribution of approximate compilations. The statistical properties we will demand of the compilation workflow are:
\begin{enumerate}
\item The compilations should all be close to the ideal, in the sense,
    \begin{align}
	\normf{U_i-V} \leq \epsilon, \quad \forall i		
		\label{eq:ass1}
	\end{align}
	\item The compilations have bounded variance, in the sense, $\E{U_i} \equiv \E{U}$ is independent of $i$, and
	\begin{align}
		\E{\normf{U_i - \E{U}}^2} \leq \epsilon', \quad \forall i
		\label{eq:ass2}
	\end{align}
\end{enumerate}
The expectations in these expressions and in the following are over the distribution of compilations generated by the compiler that satisfy Conditions 1 and 2, \ie
$ \E{f(U)} = \int {\rm d}\mu_V(U) f(U), $
where ${\rm d}\mu_V(U)$ is the measure over $SU(2^n)$ defined by the compiler output that satisfies Conditions 1 and 2 when the target is $V$ (the expectation symbol $\E{}$ should have a subscript $V$ but we omit this for ease).

First, we present a restatement of a well-known lemma.

\noindent \emph{\bf Lemma 1:} \textit{\textbf{(Mixing Lemma, Campbell \& Hastings)}} \emph{Let $V$ and $U_i$ be defined as above. Let there be probabilities $\{p_i\}_{i=1}^M, (\sum_{i=1}^M p_i=1)$, such that $\normf{\sum_{i=1}^M p_i U_i - V} \leq O(\epsilon^2)$. Then, $d_{\diamond}(\mathcal{U}, \mathcal{V}) = O(\epsilon^2)$, where $\mathcal{V}[\rho] = V \rho V^{\dagger}$ is the action of the target unitary, and $\mathcal{U}[\rho] = \sum_{i=1}^M p_i U_i \rho U_i\dg$ is the channel defined by averaging over the approximate compilations according to weights $p_i$.}

This result follows immediately from the \emph{Mixing lemma} established by Campbell \cite{PhysRevA.95.042306} and Hastings \cite{hastings2016turning}, and the observation that the Frobenius norm upper bounds the spectral norm, and therefore $\normf{\sum_{i=1}^M p_i U_i - V} \leq O(\epsilon^2) \implies \normo{\sum_{i=1}^M p_i U_i - V} \leq O(\epsilon^2)$.

It is important that the condition in Lemma 1, referred to as the \emph{reduced weighted ensemble error} (ReWEE) condition in the main text, is stated in terms of the Frobenius norm, which is easily computable with simple matrix operations. This allows us to efficiently verify it, and moreover, to specify an efficient method for determining the optimal weights for the ensemble channel $\mathcal{U}$. To see this, we explicitly write out the quantity to minimize
\begin{align}
    & \min_{\{p_i\}}~ \normf{\sum_{i=1}^{M} p_i U_i - V}^2 \nn \\
    &= \min_{\{p_i\}}~ \tr\left((\sum_{i=1}^{M} p_i U_i - V)\dg (\sum_i p_i U_i - V)\right) \nn \\
    &= \min_{\{p_i\}}~ \sum_{i,j=1}^{M} p_i p_j \tr(U_i\dg U_j) - \sum_{i=1}^{M} p_i 2\Re\tr(U_i\dg V) \nn \\
    &= \min_{\p}~ \frac{1}{2} \p^{\sf T} H \p + f^{\sf T} \p, \qquad \text{s.t.}\quad  p_i\geq 0, ~~ E^{\sf T} \p = 1 \label{eq:quadprog}
\end{align}
where $\p = [p_1, p_2, ..., p_{M}]^{\sf T}, H_{ij} = 2\Re \tr(U_i\dg U_j)$, $f_i = -2\Re\tr(U_i\dg V)$, and $E = [1, 1, ..., 1]^{\sf T}$. In the last line we have written the minimization as a standard quadratic program with inequality constraints. The matrix $H$ is a Gram matrix (since $\tr(A\dg B)$ is an inner product) and thus, positive semidefinite. This, together with the fact that the constraints define a convex set, implies that the quadratic program is convex and therefore very tractable. 

While the ReWEE condition is operationally important, we can gain more insight into the properties required by the compiler producing $U_i$ by deriving an alternative \emph{sufficient} condition on the expected properties of the compiler output distribution. To do so, first we observe that Lemma 1 holds in expectation as well. That is, if there exist probabilities $\{p_i\}_{i=1}^M, (\sum_{i=1}^M p_i=1)$, such that $\E{\normf{\sum_{i=1}^M p_i U_i - V}} \leq O(\epsilon^2)$, then $\E{d_{\diamond}(\mathcal{U}, \mathcal{V})} = O(\epsilon^2)$.

\noindent \emph{\bf Lemma 2:} \emph{Let $V$ and $U_i$ be defined as above. Given probabilities $\{p_i\}_{i=1}^M$, ($\sum_{i=1}^M p_i=1$), a sufficient condition for achieving $\E{\normf{\sum_{i=1}^M p_i U_i - V}} \leq O(\epsilon^2)$ is the \emph{reduced bias condition}: $\normf{\E{U}-V} \leq O(\epsilon^2)$.}

The proof of this Lemma, which follows from an application of the \emph{bias-variance-covariance} decomposition, a well-known decomposition of generalization error of ensembles of estimators in machine learning \cite{Ueda_1996, Brown_20005}, is provided in the SI, Section 4.

Lemma 2 provides the key property that guides the development of the compiler workflow described in the next section.  Quadratic reduction in diamond distance of an ensemble of compilations to the target circuit can be achieved if the bias of the compiler output can be quadratically reduced. Reduction of this bias is the overarching motivation for the steps in our workflow.

As the proof of Lemma 2 makes clear, if the reduced bias condition is satisfied, then one can achieve the quadratically reduced diamond distance error even with a uniform ensembles -- \ie the optimization of weights $p_i$ using the quadratic program above is not necessary, and one can simply use a uniform ensemble over compiler outputs $U_i$. However, we can get better performance by using a weighted ensemble. Even if the reduced bias condition is met, determining weights through the quadratic program allows for smaller ensemble sizes (typically much smaller) than the ensemble size bound specified by Lemma 2, $M \geq \epsilon'/\epsilon^4$ (see proof in SI, Section 4).

Lemmas 1 and 2 motivate the formation of ensemble channels constructed from approximate compilations with bounded error $\epsilon$ to achieve an output error that scales as $O(\epsilon^2)$, \emph{if} the compiler output satisfies the ReWEE condition or the reduced bias condition. Note that these conditions can be efficiently checked because they are stated in terms of an easily computable norm (Frobenius), whereas the resulting guarantee is in terms of the diamond norm. Similarly, the quadratic program that makes the ensemble efficient (in terms of number of unitaries that compose the channel) is tractable due to its convexity and the simple matrix computations required to form $H$ and $f$ in Eq. \eqref{eq:quadprog}.

Our final result establishes convergence criteria for implementing an ensemble quantum channel defined by $\{U_i, p_i\}_{i=1}^M$. 

\noindent \emph{\bf Lemma 3:} \emph{Let $V$ and $U_i$ be defined as above, and let these unitaries act on a register of $n$ qubits; \ie ${\rm dim}~ V = {\rm dim}~ U_i = 2^n \equiv d$. Define $\mathcal{V}[\rho] = V \rho V^{\dagger}$ as a target channel, and $\mathcal{U}[\rho] = \sum_{i=1}^M p_i U_i \rho U_i\dg$ as an optimized ensemble channel, such that $d_\diamond(\mathcal{V}, \mathcal{U})\leq O(\epsilon^2)$. Further, let $\hat{\mathcal{U}}[\rho] = \frac{1}{T}\sum_{i=1}^T U_{\sigma(i)} \rho U_{\sigma(i)}\dg$, with $\sigma(i)\in \{1,...,M\}$, be the empirical channel defined by sampling and averaging over $U_i$ according to the distribution $\{p_i\}$. For $\epsilon, \delta \in (0,1)$, $\pr\left\{ d_\diamond(\mathcal{V}, \hat{\mathcal{U}}) < O(\epsilon^2) \right\} > 1-\delta$, given
\begin{align}
    T \geq \frac{2v + \frac{2}{3}R \left(\epsilon^2/(2d^2)\right) }{\left(\epsilon^2/(2d^2)\right)^2}\log \frac{2d^2}{\delta}
    \label{eq:T_bound}
\end{align}
Here, $v \equiv \normo{\sum_{i=1}^M p_i (J_i - J_{\mathcal{U}})^2}, R \equiv \max_i \normo{J_i - J_{\mathcal{U}}}$, and $J_i (and J_{\mathcal{U}}$) are the $d^2$-dimensional Choi matrices associated to the channel $\mathcal{U}_i[\rho] = U_i \rho U_i^{\dagger}$ (and $\mathcal{U}$).}

The proof of this result is provided in the SI, Section 4, and is an application of the Bernstein matrix concentration inequality to an upper bound on the diamond norm provided by the difference in Choi matrices \cite{watrous_2018}. This lower bound on the number of samples required for convergence, Eq. \eqref{eq:T_bound}, strictly scales as $O(1/\epsilon^4)$, but in practice we have found that this is a vast overestimate because $v = O(\epsilon^2)$, and hence $T \geq O(1/\epsilon^2)$ suffices. The much faster convergence is consistent with the data shown in Fig. \ref{fig:td_convergences}, where we show convergence in trace distance of channel outputs.

\subsection*{Compiler workflow}
\begin{figure}[ht!]
    \centering
    \includegraphics[width=\linewidth]{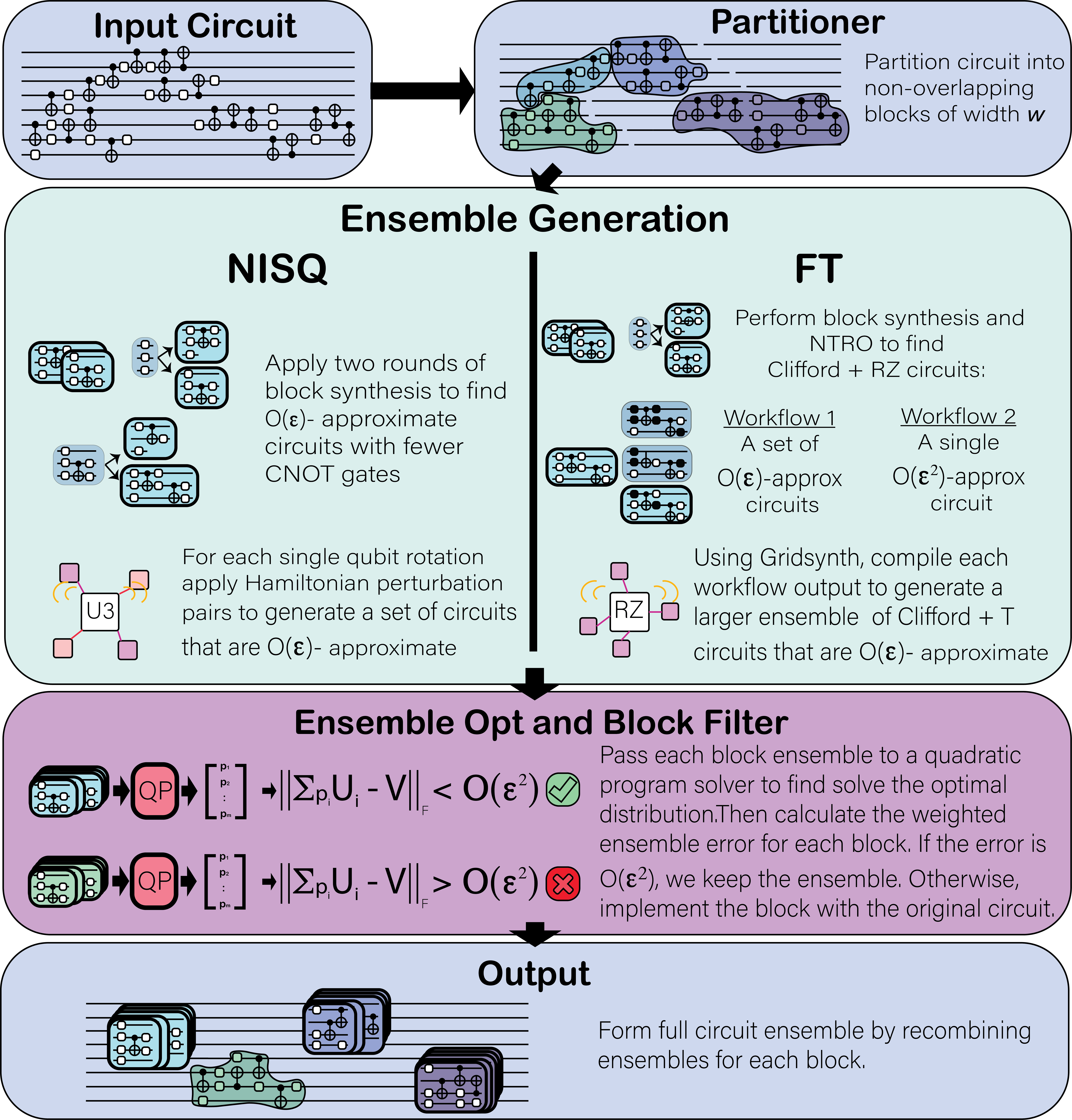}
    \caption{Detailed overview of our compilation workflow. The input circuit is partitioned into blocks of width 4. At this circuit width, we can leverage numerical optimization techniques\cite{bqskit, ntro} to find more efficient circuit representations for our inputted circuit. We diversify this initial set of solutions and pass this ensemble to a quadratic program. The quadratic program gives us the corresponding sampling distribution, with which we calculate the \emph{weighted ensemble error} of each block ensemble. We can then select those blocks which quadratically reduce the error, and recombine them back into a full circuit ensemble.}
    \label{fig:schematic}
\end{figure}

In this section we describe our full end-to-end compiler workflow, and its implementation for the demonstrations presented in the main text. Much of the implementation relies on tools integrated into BQSKit \cite{bqskit}, an open-source numerical optimization based compiler.
We can comfortably scale to algorithms with hundreds of qubits, and target both NISQ and FT gate sets. Figure \ref{fig:schematic} shows a detailed depiction of the workflow, which was outlined in Figure \ref{fig:resut_workflow} of the main text.

\subsubsection*{Circuit partitioning}
The first step in the workflow is to partition the input circuit into non-overlapping sub-circuits acting on $w$ qubits, referred to as \emph{blocks} of width $w$. In BQSKit, there are 2 partitioners which trade off quality (minimal number of partitions) for speed. The \emph{ScanPartitioner} works by iteratively selecting the best block of $w$ qubits from left to right in a circuit. After scoring all possible $w$-qubit blocks starting from the current frontier (which divides partitioned from un-partitioned sections), it picks the highest-scoring block and moves the frontier past it. This process repeats until the entire circuit is partitioned. On the other hand, the \emph{QuickPartitioner} avoids this exhaustive search of all possible blocks, and greedily groups gates that appear in topological order. For our implementation, we utilize the ScanPartitioner for smaller circuits, and rely on a combination of both partitioners for larger circuits. For these circuits, we first divide into larger blocks with the QuickPartitioner, and split those larger blocks into blocks of width $w$ with the ScanPartitioner.

The choice of block width, $w$, presents a tradeoff. Typically, the larger the block width, the more options there are for optimizing resources (\eg reducing expensive gates), and synthesizing many diverse circuit approximations of the block in the next step. However, the computational cost of approximate synthesis increases with $w$, and therefore this variable should be chosen based on computational resources and available time. In all the demonstrations in this paper, we choose $w=4$. For smaller block widths, we found that the solutions generated were rarely diverse enough to quadratically reduce the weighted ensemble error for the output ensemble. 

\subsubsection*{Ensemble Generation}
Our Ensemble Generation step of the workflow generates a set of circuits that approximate the target circuit for a block to Frobenius error $\epsilon$. The process is split into two main components: Numerical Synthesis and Diversification. Both steps are performed independently for each block and are therefore trivially parallelized. 

This step has an important distinction between the NISQ and FT workflows. When targeting NISQ architectures, Numerical Synthesis generates a single set of circuits in the CNOT + $U3$ gateset. Each member of this set of circuits satisfies $\normf{U_i-V}\leq \epsilon$. When targeting FT architectures, Numerical Synthesis generates two parallel outputs. The first output matches that of the NISQ workflow: a set of circuits in the CNOT + $Rz$ gateset that satisfies $\normf{U_i-V}\leq \epsilon$.  The second output is a \emph{single circuit} in the CNOT + $Rz$ gateset that satisfies the tighter threshold $\normf{U_i-V}\leq \epsilon ^2$. 

As we describe in further detail in the Diversification section, the reason for this is to maximize our resource reduction and probability of satisfying the ReWEE condition. The first output can achieve better resource reduction, as it gives a (quadratically) larger error budget to the Numerical Synthesis step. The second output still gives us resource reduction, but the ability to satisfy the ReWEE condition is much improved. By considering both options, we are able to choose the ensemble with the fewest resources that also quadratically reduces the error, and thereby improve the quality of our final output. Importantly, both workflows generate the same output at the end of the Diversification step: An ensemble of Clifford + T circuits that satisfies $\normf{U_i-V}\leq \epsilon$.

\emph{Numerical Synthesis}: The Numerical Synthesis step of the workflow constructs circuits with continuous angles that have \emph{fewer} resources than our target circuit. Our implementation utilizes the synthesis tools in BQSKit \cite{bqskit} for this step. Given a target unitary $V$ and native gate set $\mathcal{G}$, BQSKit constructs a family of circuits $U_i$ composed of gates in $\mathcal{G}$ that approximate $V$ by using a heuristic-guided search \cite{leap, qsearch} to generate different parametrized circuit ansatze. Then, it leverages numerical optimization to \emph{instantiate} the ansatze with the optimized parameters that minimize the Frobenius error to the target unitary, $\normf{U_i-V}$. The search step can be biased towards prioritizing circuits that minimize expensive resources such as CNOT gates or arbitrary angle rotations. In our implementation, we terminate the search and optimization as soon as the circuit ansatze contain as many expensive gates as the original target circuit, at which point we output all circuits found satisfying $\normf{U_i-V}\leq \epsilon$. While the universal gate set that each circuit is synthesized to is arbitrary, in the demonstrations presented in this paper, we synthesized to the CNOT+$U3$ gateset for NISQ implementations and the Clifford+$R_z$ gateset for FT implementations. 

For synthesis of blocks for FT implementations, on top of the heuristic-guided search, each circuit output by BQSKit undergoes another pass through the \emph{Numerical T gate Reduction} pass (NTRO) tool ~\cite{ntro}. NTRO optimizes the circuit by reducing the number of non-Clifford elements, which maintaining approximation error. NTRO reframes the instantiation problem as the constrained numerical optimization problem
\begin{equation}
    \min_\theta ~ \mathcal{C}_{\rm non-Clifford}(U_i(\theta)), \quad \text{ s.t. } \quad \|U_i(\theta) - V\|_F \leq \epsilon, \nn
\end{equation} 
where we write the synthesized circuit as $U_i(\theta)$ to explicitly indicate the dependence on the $R_z$ rotation angles within the circuit. The cost function $\mathcal{C}_{\rm non-Clifford}(U_i(\theta))$ is a heuristic that estimates the number of non-Clifford gates needed approximate unitary $V$ to precision $\epsilon$. NTRO minimizes this cost function by applying strategies that attempt to approximate each $R_z(\theta)$ in the circuit with a Clifford, or close to Clifford, rotation.

As mentioned above, in the FT case, there are two parallel outputs for Numerical Synthesis. The first output BQSKit and NTRO are executed multiple times to generate an ensemble of circuit satisfying $\|U_i(\theta) - V\|_F \leq \epsilon$. For the second output the same tools are applied to generate \emph{one} circuit, satisfying $\|U_i(\theta) - V\|_F \leq \epsilon^2$.

\emph{Diversification}: The next step in our workflow applies a number of heuristics to attempt to increase the diversity of the approximation ensembles for each block output by the Numerical Synthesis step. The aim of these heuristics is to generate an ensemble of approximation circuits that satisfy the ReWEE condition, \ie $\normf{\sum_i p_i U_i - V}\leq O(\epsilon^2)$. 

In the NISQ compilation case, each approximate solution output by Numerical Synthesis consists of CNOTs and $U3$ single qubit rotations. Since we do not want to increase the number of CNOTs our diversification algorithm only modifies the $U3$ gates. Our heuristic is motivated by the observation that for each approximate circuit $U_i$, with Frobenius error, $\normf{U_i - V} = e_i \leq \epsilon$, we have an excess error budget $\epsilon - e_i$. Our goal is to ``spend'' this remaining error budget to spread apart the initial set of solutions. A description of our heuristic solution is the following:
\begin{enumerate}
    \item Let the number of $U3$ rotations in the circuit be $N_u$. We can define the error budget per $U3$ gate as $\epsilon_{u3} = \frac{\epsilon - e_i}{N_u}$
    \item For each $U3$, generate a set of perturbations of distance $O(\epsilon_{u3})$ by generating a set of perturbing Hamiltonians, $\mathcal{P}$, that take the form $P_{abc} = aX + bY + cZ$ where $||[a,b,c]||_2 = \epsilon_{u3}$. Importantly, each time an operator with coefficients $[a,b,c]$ is added to the set, one with the opposing coefficient $[-a,-b,-c]$ must also be added to ensure local convexity. The total number of perturbations generated is user-specified.
    \item Generate a set of perturbed unitaries for each $U3$ by post-multiplication: $U3 \cdot e^{i P_{abc}}$, for all perturbating Hamiltonians $P_{abc} \in \mathcal{P}$. We then rewrite each perturbed single-qubit unitary as an explicit $U3$ rotation.
    \item Finally, we confirm that $\normf{U_i-V} \leq \epsilon$ for each perturbed version of $U_i$.
\end{enumerate}

In the FT compilation case, each approximate solution output by the Numerical Synthesis step consists of Clifford and $R_z$ gates. Each $R_z$ rotation needs to be approximated by a set of Clifford+$T$ single qubit rotations, and the non-deterministic procedure for doing this also allows us to introduce diversity. While we can perturb each $R_z$ angle by the error budget to expand the circuit, Campbell \cite{PhysRevA.95.042306} presents a method to generate a set of $O(\epsilon)$ approximations for an axial rotation with a resulting error of $O(\epsilon^2)$. The method depends upon an oracle to find an $O(\epsilon)$ approximation, for which we use Ross and Sellinger's Gridsynth\cite{ross_sellinger} algorithm. Gridsynth outputs a series of Clifford and $T$ gates that approximate an angle to a given precision. This allows us to perform an $Rz$ gate fault-tolerantly without additional ancilla qubits. It is shown in Kuchlinov \emph{et al.} \cite{Kliuchnikov_2023} that we can expand this methodology to ancilla-based techniques as well, which may save even more $T$ gates. This method not only diversifies our set of solutions, but removes additional $T$ gates in our final ensemble. At the end of this step, we again confirm that $\normf{U_i-V} \leq \epsilon$ for each perturbed version of $U_i$.

For the FT workflow, Diversification converts the output of synthesis to the target gate set, Clifford+$T$. Additionally, we are able to able to reduce the number of $T$ gates by compiling each $Rz$ to $O(\epsilon)$. Because of this, we are able to able to consider an additional workflow with a quadratically smaller budget passed to Numerical Synthesis ($\epsilon^2$) and a Diversification budget of $O(\epsilon)$. This ensemble has a much higher probability of satisfying the ReWEE condition, since we diversify each $Rz$ gate in a way that guarantees quadratic error reduction locally (with respect to the single qubit unitary). By considering a single circuit that has a distance $\epsilon^2$, the resultant ensemble is likely to form a channel with $O(\epsilon^2)$ distance from the target channel. This increased probability of meeting the ReWEE condition in the FT setting is reflected in the statistics of weighted ensemble error shown in SI, Section 3 (bottom plot of Figure 2). As mentioned, the trade off of this second workflow is that we limit the resource savings attained from the Numerical Synthesis step. By considering both outputs, we can choose the ensemble that saves the most resources and quadratically reduces the error.

In both the NISQ and FT cases, at the end of the diversification step, we end up with a larger ensemble for each block of partitioned circuit, with each member of the ensemble having $\epsilon$ error to the target unitary for that block.

\subsubsection*{Filtering out Blocks}
\label{sec:bias_filter}
The preceding Ensemble Generation steps are designed to construct ensembles of approximate circuits for each block of the circuit that satisfy the ReWEE condition. While this property is not guaranteed by these steps we can explicitly check it for each block. It is important to note that this verification is possible because (i) the optimization required to formulate the optimal weighted ensemble, and (ii) the ReWEE condition itself, are both efficiently computable. 

Therefore, in this step we perform the convex quadratic optimization over the ensemble for each block, and compute the weighted ensemble error, $\normf{\sum_{i=1}^{M^\k}p^\k_i U^\k_i - V^\k}$. If this quantity is $\leq C\epsilon^2$ we store this weighted ensemble for block $k$. In the demonstrations shown in the main text we chose $C=2$ for $\epsilon=10^{-1}$, and $C=20$ for $\epsilon \leq 10^{-2}$. If this error bound is not satisfied, we revert back to the original target circuit for the block, and the size of the ``ensemble'' for that block is one. In the FT workflow, we only revert back to the original circuit if both output ensembles do not satisfy the error bound. We refer to such blocks where the REWEE condition cannot be met while simultaneously reducing resources as ``rigid'' blocks. Intuitively, rigid blocks contain mostly fixed-angle gates, many of which are in the set of expensive gates. For example, if a block only contains a sequence of CNOT gates, it is impossible to reduce the number of CNOTs while also maintaining the ReWEE condition for $\epsilon$ below some value. In the SI we provide statistics that summarize the number of blocks that satisfy the ReWEE condition and the sufficient reduced bias condition for each of our benchmarks. 

$C$ is a hyper-parameter in this implementation of the workflow. Small values of $C$ provide tighter guarantees on the output error, while larger values allow for more resource savings. Note that even for large values of $C$, the overall output error can be bounded by $K\epsilon^2$ because even if some blocks have error close to $C\epsilon^2$, others can have much smaller error.

Note that there exist synthesis algorithms that guarantee that the ReWEE condition is met, \eg Campbell's convex hull algorithm \cite{PhysRevA.95.042306}. However, these provide no guarantees on resource usage and lead to extravagant use of expensive gates in order to satisfy the ReWEE condition for rigid blocks. Our technique maximizes generality and resource optimization while remaining scalable in order to be applied off-the-shelf to any input circuit. As a tradeoff, resource reduction is not possible for all blocks.

\subsubsection*{Generating the Full Circuit Ensemble}
The final step of our workflow is to combine all of the block ensembles into an ensemble over the entire circuit. 
Given optimized ensembles for $K$ partitions, this defines a joint distribution $\mathbb{P} \equiv \mathbb{P}_1 \times ... \times \mathbb{P}_K$, from which circuits are sampled when executing the full-scale circuit.

\section*{Acknowledgments}

This work was supported by the U.S. Department of Energy, Office of Science, Office of Advanced Scientific Computing Research through the Accelerated Research in Quantum Computing Program MACH-Q Project. This research used resources of the National Energy Research Scientific Computing Center (NERSC), a Department of Energy Office of Science User Facility using NERSC award DDR-ERCAPm4141. M.S. also acknowledges support from the Laboratory
Directed Research and Development program (Project 233972) at Sandia National Laboratories. Sandia National Laboratories is a multimission laboratory managed and operated by National Technology and Engineering Solutions of Sandia LLC, a wholly owned subsidiary of Honeywell International Inc. for the U.S. Department of Energy’s National Nuclear Security Administration under contract DE-NA0003525.


\begin{thebibliography}{100}

\bibitem{reiher_femoco_2017}
Markus Reiher, Nathan Wiebe, Krysta~M. Svore, Dave Wecker, and Matthias Troyer.
\newblock Elucidating reaction mechanisms on quantum computers.
\newblock {\em Proceedings of the National Academy of Sciences},
  114(29):7555--7560, 2017.

\bibitem{vonburg_2021}
Vera von Burg, Guang~Hao Low, Thomas H\"aner, Damian~S. Steiger, Markus Reiher,
  Martin Roetteler, and Matthias Troyer.
\newblock Quantum computing enhanced computational catalysis.
\newblock {\em Phys. Rev. Res.}, 3:033055, Jul 2021.

\bibitem{PRXQuantum.2.010103}
Ryan Babbush, Jarrod~R. McClean, Michael Newman, Craig Gidney, Sergio Boixo,
  and Hartmut Neven.
\newblock Focus beyond quadratic speedups for error-corrected quantum
  advantage.
\newblock {\em PRX Quantum}, 2:010103, Mar 2021.

\bibitem{su_2021}
Yuan Su, Dominic~W. Berry, Nathan Wiebe, Nicholas Rubin, and Ryan Babbush.
\newblock Fault-tolerant quantum simulations of chemistry in first
  quantization.
\newblock {\em PRX Quantum}, 2:040332, Nov 2021.

\bibitem{ShokrianZini2023}
Modjtaba Shokrian~Zini, Alain Delgado, Roberto dos Reis, Pablo~Antonio
  Moreno~Casares, Jonathan~E. Mueller, Arne-Christian Voigt, and Juan~Miguel
  Arrazola.
\newblock Quantum simulation of battery materials using ionic pseudopotentials.
\newblock {\em {Quantum}}, 7:1049, July 2023.

\bibitem{rubin_2024}
Nicholas~C. Rubin, Dominic~W. Berry, Alina Kononov, Fionn~D. Malone, Tanuj
  Khattar, Alec White, Joonho Lee, Hartmut Neven, Ryan Babbush, and Andrew~D.
  Baczewski.
\newblock Quantum computation of stopping power for inertial fusion target
  design.
\newblock {\em Proceedings of the National Academy of Sciences},
  121(23):e2317772121, 2024.

\bibitem{PRXQuantum.5.020101}
Amara Katabarwa, Katerina Gratsea, Athena Caesura, and Peter~D. Johnson.
\newblock Early fault-tolerant quantum computing.
\newblock {\em PRX Quantum}, 5:020101, Jun 2024.

\bibitem{Preskill2018_NISQ}
John Preskill.
\newblock Quantum {C}omputing in the {NISQ} era and beyond.
\newblock {\em {Quantum}}, 2:79, August 2018.

\bibitem{Kim2023}
Youngseok Kim, Andrew Eddins, Sajant Anand, Ken~Xuan Wei, Ewout van~den Berg,
  Sami Rosenblatt, Hasan Nayfeh, Yantao Wu, Michael Zaletel, Kristan Temme, and
  Abhinav Kandala.
\newblock Evidence for the utility of quantum computing before fault tolerance.
\newblock {\em Nature}, 618(7965):500--505, Jun 2023.

\bibitem{botea2018complexity}
Adi Botea, Akihiro Kishimoto, and Radu Marinescu.
\newblock On the complexity of quantum circuit compilation.
\newblock In {\em Proceedings of the International Symposium on Combinatorial
  Search}, volume~9, pages 138--142, 2018.

\bibitem{wu_qgo}
Xin-Chuan Wu, Marc~Grau Davis, Frederic~T. Chong, and Costin Iancu.
\newblock Reoptimization of quantum circuits via hierarchical synthesis.
\newblock In {\em 2021 International Conference on Rebooting Computing (ICRC)},
  pages 35--46, 2021.

\bibitem{PhysRevA.95.042306}
Earl Campbell.
\newblock Shorter gate sequences for quantum computing by mixing unitaries.
\newblock {\em Phys. Rev. A}, 95:042306, Apr 2017.

\bibitem{hastings2016turning}
Matthew~B Hastings.
\newblock Turning gate synthesis errors into incoherent errors.
\newblock {\em arXiv:1612.01011}, 2016.

\bibitem{martyn2024}
John~M. Martyn and Patrick Rall.
\newblock Halving the cost of quantum algorithms with randomization, 2024.

\bibitem{campbell_qdrift}
Earl Campbell.
\newblock Random compiler for fast hamiltonian simulation.
\newblock {\em Phys. Rev. Lett.}, 123:070503, Aug 2019.

\bibitem{Akibue2024}
Seiseki Akibue, Go~Kato, and Seiichiro Tani.
\newblock Probabilistic state synthesis based on optimal convex approximation.
\newblock {\em npj Quantum Information}, 10:3, Jan 2024.

\bibitem{yoshioka2024errorcrafting}
Nobuyuki Yoshioka, Seiseki Akibue, Hayata Morisaki, Kento Tsubouchi, and
  Yasunari Suzuki.
\newblock Error crafting in probabilistic quantum gate synthesis, 2024.

\bibitem{Akibu_ACM_2024}
Seiseki Akibue, Go~Kato, and Seiichiro Tani.
\newblock Probabilistic unitary synthesis with optimal accuracy.
\newblock {\em ACM Transactions on Quantum Computing}, 5, 2024.

\bibitem{Ouyang2020compilation}
Yingkai Ouyang, David~R. White, and Earl~T. Campbell.
\newblock Compilation by stochastic {H}amiltonian sparsification.
\newblock {\em {Quantum}}, 4:235, February 2020.

\bibitem{PRXQuantum.2.040305}
Chi-Fang Chen, Hsin-Yuan Huang, Richard Kueng, and Joel~A. Tropp.
\newblock Concentration for random product formulas.
\newblock {\em PRX Quantum}, 2:040305, Oct 2021.

\bibitem{PhysRevResearch.6.013224}
Matthew Pocrnic, Matthew Hagan, Juan Carrasquilla, Dvira Segal, and Nathan
  Wiebe.
\newblock Composite qdrift-product formulas for quantum and classical
  simulations in real and imaginary time.
\newblock {\em Phys. Rev. Res.}, 6:013224, Mar 2024.

\bibitem{huang_2023}
Yifei Huang, Yuguo Shao, Weiluo Ren, Jinzhao Sun, and Dingshun Lv.
\newblock Efficient quantum imaginary time evolution by drifting real-time
  evolution: An approach with low gate and measurement complexity.
\newblock {\em Journal of Chemical Theory and Computation}, 19(13):3868--3876,
  2023.

\bibitem{PhysRevA.109.062431}
Chien-Hung Cho, Dominic~W. Berry, and Min-Hsiu Hsieh.
\newblock Doubling the order of approximation via the randomized product
  formula.
\newblock {\em Phys. Rev. A}, 109:062431, Jun 2024.

\bibitem{patel_2021}
Tirthak Patel, Ed~Younis, Costin Iancu, Wibe de~Jong, and Devesh Tiwari.
\newblock Robust and resource-efficient quantum circuit approximation, 2021.

\bibitem{low2021halving}
Guang~Hao Low.
\newblock Halving the cost of quantum multiplexed rotations, 2021.

\bibitem{Kliuchnikov_2023}
Vadym Kliuchnikov, Kristin Lauter, Romy Minko, Adam Paetznick, and Christophe
  Petit.
\newblock Shorter quantum circuits via single-qubit gate approximation.
\newblock {\em {Quantum}}, 7:1208, December 2023.

\bibitem{Burt_2024}
Felix Burt, Kuan-Cheng Chen, and Kin~K. Leung.
\newblock Generalised circuit partitioning for distributed quantum computing.
\newblock In {\em 2024 IEEE International Conference on Quantum Computing and
  Engineering (QCE)}, page 173–178. IEEE, September 2024.

\bibitem{Ueda_1996}
N.~Ueda and R.~Nakano.
\newblock Generalization error of ensemble estimators.
\newblock In {\em Proceedings of International Conference on Neural Networks
  (ICNN'96)}, volume~1, pages 90--95 vol.1, 1996.

\bibitem{Brown_20005}
Gavin Brown, Jeremy~L. Wyatt, and Peter Ti{\~n}o.
\newblock Managing diversity in regression ensembles.
\newblock {\em Journal of Machine Learning Research}, 6(55):1621--1650, 2005.

\bibitem{tropp_2015}
Joel~A. Tropp.
\newblock An introduction to matrix concentration inequalities.
\newblock {\em Found. Trends Mach. Learn.}, 8(1–2):1–230, May 2015.

\bibitem{watrous_2018}
J.~Watrous.
\newblock {\em The Theory of Quantum Information}.
\newblock Cambridge University Press, 2018.

\bibitem{noauthor_notitle_nodate}


\bibitem{noauthor_notitle_nodate-1}


\bibitem{hennessy_computer_2012}
John~L. Hennessy, David~A. Patterson, and Krste Asanović.
\newblock {\em Computer architecture: a quantitative approach}.
\newblock Morgan Kaufmann/Elsevier, Waltham, MA, 5th ed edition, 2012.
\newblock OCLC: ocn755102367.

\bibitem{carbone_controlled_1970}
G.~Carbone and L.~Massone.
\newblock [{Controlled} experimentation on the therapeutic effect of 2 topical
  anti-inflammatory steroids].
\newblock {\em Archivio Italiano Di Dermatologia, Venereologia, E Sessuologia},
  36(3):207--214, 1970.

\bibitem{noauthor_notitle_nodate-2}


\bibitem{ieee_computer_society_ispass_2009}
IEEE~Computer Society, editor.
\newblock {\em {ISPASS} 2009: {IEEE} {International} {Symposium} on
  {Performance} {Analaysis} of {Systems} and {Software}: {April} 26-28, 2009,
  {Boston}, {MA}, {USA}}.
\newblock Institute of Electrical and Electronics Engineers, Piscataway, NJ,
  2009.
\newblock Meeting Name: International Symposium on Performance Analysis of
  Systems and Software.

\bibitem{noauthor_2012_2012}
{\em 2012 innovative parallel computing ({InPar} 2012): {San} {Jose},
  {California}, {USA}, 13 - 14 {May} 2012}.
\newblock IEEE, Piscataway, NJ, 2012.

\bibitem{gupta_study_2012}
Kshitij Gupta, Jeff~A. Stuart, and John~D. Owens.
\newblock A study of {Persistent} {Threads} style {GPU} programming for {GPGPU}
  workloads.
\newblock In {\em 2012 {Innovative} {Parallel} {Computing} ({InPar})}, pages
  1--14, San Jose, CA, USA, May 2012. IEEE.

\bibitem{laine_megakernels_2013}
Samuli Laine, Tero Karras, and Timo Aila.
\newblock Megakernels considered harmful: wavefront path tracing on {GPUs}.
\newblock In {\em Proceedings of the 5th {High}-{Performance} {Graphics}
  {Conference} on - {HPG} '13}, page 137, Anaheim, California, 2013. ACM Press.

\bibitem{chen_effisha_2017}
Guoyang Chen, Yue Zhao, Xipeng Shen, and Huiyang Zhou.
\newblock {EffiSha}: {A} {Software} {Framework} for {Enabling} {Effficient}
  {Preemptive} {Scheduling} of {GPU}.
\newblock In {\em Proceedings of the 22nd {ACM} {SIGPLAN} {Symposium} on
  {Principles} and {Practice} of {Parallel} {Programming}}, pages 3--16, Austin
  Texas USA, January 2017. ACM.

\bibitem{gupta_study_2012-1}
Kshitij Gupta, Jeff~A. Stuart, and John~D. Owens.
\newblock A study of {Persistent} {Threads} style {GPU} programming for {GPGPU}
  workloads.
\newblock In {\em 2012 {Innovative} {Parallel} {Computing} ({InPar})}, pages
  1--14, San Jose, CA, USA, May 2012. IEEE.

\bibitem{wu_enabling_2015}
Bo~Wu, Guoyang Chen, Dong Li, Xipeng Shen, and Jeffrey Vetter.
\newblock Enabling and {Exploiting} {Flexible} {Task} {Assignment} on {GPU}
  through {SM}-{Centric} {Program} {Transformations}.
\newblock In {\em Proceedings of the 29th {ACM} on {International} {Conference}
  on {Supercomputing}}, pages 119--130, Newport Beach California USA, June
  2015. ACM.

\bibitem{plano_scheduling_2020}
Tom Plano and Jeremy Buhler.
\newblock Scheduling {Irregular} {Dataflow} {Pipelines} on {SIMD}
  {Architectures}.
\newblock In {\em Proceedings of the 2020 {Sixth} {Workshop} on {Programming}
  {Models} for {SIMD}/{Vector} {Processing}}, pages 1--9, San Diego CA USA,
  February 2020. ACM.

\bibitem{ren_extracting_2019}
Bin Ren, Shruthi Balakrishna, Youngjoon Jo, Sriram Krishnamoorthy, Kunal
  Agrawal, and Milind Kulkarni.
\newblock Extracting {SIMD} {Parallelism} from {Recursive} {Task}-{Parallel}
  {Programs}.
\newblock {\em ACM Transactions on Parallel Computing}, 6(4):1--37, December
  2019.

\bibitem{troendle_specialized_2019}
David Troendle, Tuan Ta, and Byunghyun Jang.
\newblock A {Specialized} {Concurrent} {Queue} for {Scheduling} {Irregular}
  {Workloads} on {GPUs}.
\newblock In {\em Proceedings of the 48th {International} {Conference} on
  {Parallel} {Processing}}, pages 1--11, Kyoto Japan, August 2019. ACM.

\bibitem{mccully_rna_1988}
J.~D. McCully and C.~C. Liew.
\newblock {RNA} transcription in myocardial-cell nuclei during postnatal
  development. {A} study establishing an assay system for transcription in
  vitro.
\newblock {\em The Biochemical Journal}, 256(2):441--445, December 1988.

\bibitem{huo_efficient_2013}
Xin Huo, Sriram Krishnamoorthy, and Gagan Agrawal.
\newblock Efficient scheduling of recursive control flow on {GPUs}.
\newblock In {\em Proceedings of the 27th international {ACM} conference on
  {International} conference on supercomputing - {ICS} '13}, page 409, Eugene,
  Oregon, USA, 2013. ACM Press.

\bibitem{noauthor_cuda_nodate}
{CUDA} {C}++ {Programming} {Guide}.
\newblock page 441.

\bibitem{anantpur_taming_2014}
Jayvant Anantpur and Govindarajan R.
\newblock Taming {Control} {Divergence} in {GPUs} through {Control} {Flow}
  {Linearization}.
\newblock In Albert Cohen, editor, {\em Compiler {Construction}}, pages
  133--153, Berlin, Heidelberg, 2014. Springer Berlin Heidelberg.

\bibitem{anantpur_taming_2014-1}
Jayvant Anantpur and Govindarajan R.
\newblock Taming {Control} {Divergence} in {GPUs} through {Control} {Flow}
  {Linearization}.
\newblock In Albert Cohen, editor, {\em Compiler {Construction}}, pages
  133--153, Berlin, Heidelberg, 2014. Springer Berlin Heidelberg.

\bibitem{huang_logical_2021}
Yipeng Huang, Steven Holtzen, Todd Millstein, Guy Van~den Broeck, and Margaret
  Martonosi.
\newblock Logical {Abstractions} for {Noisy} {Variational} {Quantum}
  {Algorithm} {Simulation}.
\newblock {\em arXiv:2103.17226 [quant-ph]}, March 2021.
\newblock arXiv: 2103.17226.

\bibitem{maccormack_simulating_2021}
Ian MacCormack, Alexey Galda, and Adam~L. Lyon.
\newblock Simulating {Large} {PEPs} {Tensor} {Networks} on {Small} {Quantum}
  {Devices}.
\newblock {\em arXiv:2110.00507 [cond-mat, physics:quant-ph]}, October 2021.
\newblock arXiv: 2110.00507.

\bibitem{markov_simulating_2008}
Igor~L. Markov and Yaoyun Shi.
\newblock Simulating {Quantum} {Computation} by {Contracting} {Tensor}
  {Networks}.
\newblock {\em SIAM Journal on Computing}, 38(3):963--981, January 2008.

\bibitem{markov_simulating_2008-1}
Igor~L. Markov and Yaoyun Shi.
\newblock Simulating {Quantum} {Computation} by {Contracting} {Tensor}
  {Networks}.
\newblock {\em SIAM Journal on Computing}, 38(3):963--981, January 2008.
\newblock Publisher: Society for Industrial and Applied Mathematics.

\bibitem{allen_quantum_2017}
John-Mark~A. Allen, Jonathan Barrett, Dominic~C. Horsman, Ciarán~M. Lee, and
  Robert~W. Spekkens.
\newblock Quantum {Common} {Causes} and {Quantum} {Causal} {Models}.
\newblock {\em Physical Review X}, 7(3):031021, July 2017.

\bibitem{allen_quantum_2017-1}
John-Mark~A. Allen, Jonathan Barrett, Dominic~C. Horsman, Ciarán~M. Lee, and
  Robert~W. Spekkens.
\newblock Quantum {Common} {Causes} and {Quantum} {Causal} {Models}.
\newblock {\em Physical Review X}, 7(3):031021, July 2017.
\newblock Publisher: American Physical Society.

\bibitem{nazarov_bayesian_2020}
Ivan Nazarov and Evgeny Burnaev.
\newblock Bayesian {Sparsification} {Methods} for {Deep} {Complex}-valued
  {Networks}.
\newblock {\em arXiv:2003.11413 [cs, stat]}, June 2020.
\newblock arXiv: 2003.11413.

\bibitem{gottesman_heisenberg_1998}
Daniel Gottesman.
\newblock The {Heisenberg} {Representation} of {Quantum} {Computers}.
\newblock {\em arXiv:quant-ph/9807006}, July 1998.
\newblock arXiv: quant-ph/9807006.

\bibitem{aaronson_improved_2004}
Scott Aaronson and Daniel Gottesman.
\newblock Improved simulation of stabilizer circuits.
\newblock {\em Physical Review A}, 70(5):052328, November 2004.
\newblock Publisher: American Physical Society.

\bibitem{emerson_scalable_2005}
Joseph Emerson, Robert Alicki, and Karol Zyczkowski.
\newblock Scalable {Noise} {Estimation} with {Random} {Unitary} {Operators}.
\newblock {\em Journal of Optics B: Quantum and Semiclassical Optics},
  7(10):S347--S352, October 2005.
\newblock arXiv: quant-ph/0503243.

\bibitem{aaronson_complexity-theoretic_2016}
Scott Aaronson and Lijie Chen.
\newblock Complexity-{Theoretic} {Foundations} of {Quantum} {Supremacy}
  {Experiments}.
\newblock {\em arXiv:1612.05903 [quant-ph]}, December 2016.
\newblock arXiv: 1612.05903.

\bibitem{bqskit}
Ed~Younis, Costin~C. Iancu, Wim Lavrijsen, Marc Davis, and Ethan Smith.
\newblock Berkeley quantum synthesis toolkit (bqskit) v1.
\newblock [Computer Software] \url{https://doi.org/10.11578/dc.20210603.2}, apr
  2021.

\bibitem{qsearch}
Marc~G. Davis, Ethan Smith, Ana Tudor, Koushik Sen, Irfan Siddiqi, and Costin
  Iancu.
\newblock Towards optimal topology aware quantum circuit synthesis.
\newblock In {\em 2020 IEEE International Conference on Quantum Computing and
  Engineering (QCE)}, pages 223--234, 2020.

\bibitem{ntro}
Marc~Grau Davis.
\newblock {\em Numerical Synthesis of Arbitrary Multi-Qubit Unitaries with low
  T-Count}.
\newblock PhD thesis, Massachusetts Institute of Technology, 02 2023.

\bibitem{leap}
Ethan Smith, Marc~Grau Davis, Jeffrey Larson, Ed~Younis, Lindsay~Bassman
  Oftelie, Wim Lavrijsen, and Costin Iancu.
\newblock Leap: Scaling numerical optimization based synthesis using an
  incremental approach.
\newblock {\em ACM Transactions on Quantum Computing}, 4(1), February 2023.

\bibitem{emerson_scalable_2005-1}
Joseph Emerson, Robert Alicki, and Karol Zyczkowski.
\newblock Scalable {Noise} {Estimation} with {Random} {Unitary} {Operators}.
\newblock {\em Journal of Optics B: Quantum and Semiclassical Optics},
  7(10):S347--S352, October 2005.
\newblock arXiv: quant-ph/0503243.

\bibitem{gilchrist_distance_2005}
Alexei Gilchrist, Nathan~K. Langford, and Michael~A. Nielsen.
\newblock Distance measures to compare real and ideal quantum processes.
\newblock {\em Physical Review A}, 71(6):062310, June 2005.
\newblock arXiv: quant-ph/0408063.

\bibitem{deutsch_quantum_1989}
David~Elieser Deutsch and Roger Penrose.
\newblock Quantum computational networks.
\newblock {\em Proceedings of the Royal Society of London. A. Mathematical and
  Physical Sciences}, 425(1868):73--90, September 1989.
\newblock Publisher: Royal Society.

\bibitem{tannu_case_2018}
Swamit~S. Tannu and Moinuddin~K. Qureshi.
\newblock A {Case} for {Variability}-{Aware} {Policies} for {NISQ}-{Era}
  {Quantum} {Computers}.
\newblock {\em arXiv:1805.10224 [quant-ph]}, May 2018.
\newblock arXiv: 1805.10224.

\bibitem{abhari_scaffold_nodate}
Ali~Javadi Abhari, Arvin Faruque, Mohammad~Javad Dousti, Lukas Svec, Oana Catu,
  Amlan Chakrabati, Chen-Fu Chiang, Seth Vanderwilt, John Black, Fred Chong,
  Margaret Martonosi, Martin Suchara, Ken Brown, Massoud Pedram, and Todd Brun.
\newblock Scaffold: {Quantum} {Programming} {Language}.
\newblock page~43.

\bibitem{qdrift}
Earl Campbell.
\newblock Random compiler for fast hamiltonian simulation.
\newblock {\em Physical Review Letters}, 123(7), August 2019.

\bibitem{sarovar_detecting_2020}
Mohan Sarovar, Timothy Proctor, Kenneth Rudinger, Kevin Young, Erik Nielsen,
  and Robin Blume-Kohout.
\newblock Detecting crosstalk errors in quantum information processors.
\newblock {\em Quantum}, 4:321, September 2020.
\newblock arXiv: 1908.09855.

\bibitem{rossi_quantum_2021}
A.~Rossi, P.~G. Baity, V.~M. Schäfer, and M.~Weides.
\newblock Quantum computing hardware in the cloud: {Should} a computational
  chemist care?
\newblock {\em International Journal of Quantum Chemistry}, 121(14), July 2021.
\newblock arXiv: 2102.03248.

\bibitem{divincenzo_physical_2000}
David~P. DiVincenzo.
\newblock The {Physical} {Implementation} of {Quantum} {Computation}.
\newblock {\em Fortschritte der Physik}, 48(9-11):771--783, 2000.
\newblock \_eprint:
  https://onlinelibrary.wiley.com/doi/pdf/10.1002/1521-3978\%28200009\%2948\%3A9/11\%3C771\%3A\%3AAID-PROP771\%3E3.0.CO\%3B2-E.

\bibitem{qiskit}
Ali Javadi-Abhari, Matthew Treinish, Kevin Krsulich, Christopher~J. Wood, Jake
  Lishman, Julien Gacon, Simon Martiel, Paul~D. Nation, Lev~S. Bishop,
  Andrew~W. Cross, Blake~R. Johnson, and Jay~M. Gambetta.
\newblock Quantum computing with {Q}iskit, 2024.

\bibitem{tomesh_supermarq_2022-1}
Teague Tomesh, Pranav Gokhale, Victory Omole, Gokul~Subramanian Ravi,
  Kaitlin~N. Smith, Joshua Viszlai, Xin-Chuan Wu, Nikos Hardavellas,
  Margaret~R. Martonosi, and Frederic~T. Chong.
\newblock {SupermarQ}: {A} {Scalable} {Quantum} {Benchmark} {Suite}.
\newblock {\em arXiv:2202.11045 [quant-ph]}, April 2022.
\newblock arXiv: 2202.11045.

\bibitem{nielsen_gate_2021}
Erik Nielsen, John~King Gamble, Kenneth Rudinger, Travis Scholten, Kevin Young,
  and Robin Blume-Kohout.
\newblock Gate {Set} {Tomography}.
\newblock {\em Quantum}, 5:557, October 2021.
\newblock arXiv: 2009.07301.

\bibitem{calderbank_good_1996}
A.~R. Calderbank and Peter~W. Shor.
\newblock Good {Quantum} {Error}-{Correcting} {Codes} {Exist}.
\newblock {\em Physical Review A}, 54(2):1098--1105, August 1996.
\newblock arXiv: quant-ph/9512032.

\bibitem{steane_tutorial_nodate}
Andrew~M Steane.
\newblock A {Tutorial} on {Quantum} {Error} {Correction}.
\newblock page~24.

\bibitem{steane_simple_1996}
Andrew Steane.
\newblock Simple {Quantum} {Error} {Correcting} {Codes}.
\newblock {\em Physical Review A}, 54(6):4741--4751, December 1996.
\newblock arXiv: quant-ph/9605021.

\bibitem{munoz-coreas_t-count_2018}
Edgard Muñoz-Coreas and Himanshu Thapliyal.
\newblock T-count and {Qubit} {Optimized} {Quantum} {Circuit} {Design} of the
  {Non}-{Restoring} {Square} {Root} {Algorithm}.
\newblock {\em arXiv:1712.08254 [quant-ph]}, October 2018.
\newblock arXiv: 1712.08254.

\bibitem{whitney_fault_2009}
Mark~G. Whitney, Nemanja Isailovic, Yatish Patel, and John Kubiatowicz.
\newblock A {Fault} {Tolerant}, {Area} {Efficient} {Architecture} for {Shor}'s
  {Factoring} {Algorithm}.
\newblock Technical Report arXiv:0909.2188, arXiv, September 2009.
\newblock arXiv:0909.2188 [quant-ph] type: article.

\bibitem{cacciapuoti_quantum_2020}
Angela~Sara Cacciapuoti, Marcello Caleffi, Francesco Tafuri, Francesco~Saverio
  Cataliotti, Stefano Gherardini, and Giuseppe Bianchi.
\newblock Quantum {Internet}: {Networking} {Challenges} in {Distributed}
  {Quantum} {Computing}.
\newblock {\em IEEE Network}, 34(1):137--143, January 2020.
\newblock Conference Name: IEEE Network.

\bibitem{lo_classical_2000}
Hoi-Kwong Lo.
\newblock Classical {Communication} {Cost} in {Distributed} {Quantum}
  {Information} {Processing} - {A} generalization of {Quantum} {Communication}
  {Complexity}.
\newblock {\em Physical Review A}, 62(1):012313, June 2000.
\newblock arXiv:quant-ph/9912009.

\bibitem{cacciapuoti_when_2020}
Angela~Sara Cacciapuoti, Marcello Caleffi, Rodney Van~Meter, and Lajos Hanzo.
\newblock When {Entanglement} meets {Classical} {Communications}: {Quantum}
  {Teleportation} for the {Quantum} {Internet} ({Invited} {Paper}).
\newblock {\em IEEE Transactions on Communications}, 68(6):3808--3833, June
  2020.
\newblock arXiv:1907.06197 [quant-ph].

\bibitem{caleffi_optimal_2017}
Marcello Caleffi.
\newblock Optimal {Routing} for {Quantum} {Networks}.
\newblock {\em IEEE Access}, 5:22299--22312, 2017.
\newblock Conference Name: IEEE Access.

\bibitem{daei_optimized_2020}
Omid Daei, Keivan Navi, and Mariam Zomorodi-Moghadam.
\newblock Optimized {Quantum} {Circuit} {Partitioning}.
\newblock {\em International Journal of Theoretical Physics},
  59(12):3804--3820, December 2020.
\newblock arXiv:2005.11614 [quant-ph].

\bibitem{tang_cutqc_2021-3}
Wei Tang, Teague Tomesh, Martin Suchara, Jeffrey Larson, and Margaret
  Martonosi.
\newblock {CutQC}: using small {Quantum} computers for large {Quantum} circuit
  evaluations.
\newblock In {\em Proceedings of the 26th {ACM} {International} {Conference} on
  {Architectural} {Support} for {Programming} {Languages} and {Operating}
  {Systems}}, pages 473--486, Virtual USA, April 2021. ACM.

\bibitem{gay_communicating_2004}
Simon Gay and Rajagopal Nagarajan.
\newblock Communicating {Quantum} {Processes}.
\newblock Technical Report arXiv:quant-ph/0409052, arXiv, September 2004.
\newblock arXiv:quant-ph/0409052 type: article.

\bibitem{nguyen_extending_2020}
Thien Nguyen, Anthony Santana, Tyler Kharazi, Daniel Claudino, Hal Finkel, and
  Alexander McCaskey.
\newblock Extending {C}++ for {Heterogeneous} {Quantum}-{Classical}
  {Computing}.
\newblock Technical Report arXiv:2010.03935, arXiv, October 2020.
\newblock arXiv:2010.03935 [quant-ph] type: article.

\bibitem{mosca_quantum_2019}
Michele Mosca, Martin Roetteler, and Peter Selinger.
\newblock Quantum {Programming} {Languages} ({Dagstuhl} {Seminar} 18381).
\newblock page 21 pages, 2019.
\newblock Artwork Size: 21 pages Medium: application/pdf Publisher: Schloss
  Dagstuhl - Leibniz-Zentrum fuer Informatik GmbH, Wadern/Saarbruecken, Germany
  Version Number: 1.0.

\bibitem{gao_faithful_2019}
Wei-Chao Gao, Si-Chen Mi, Cong Cao, Xiaofei Liu, Tie-Jun Wang, and Chuan Wang.
\newblock Faithful {Transmission} of {Single}-{Photon} {Qubits} {Using}
  {Error}-{Rejection} {Coding}.
\newblock {\em IEEE Photonics Journal}, 11(1):1--7, February 2019.
\newblock Conference Name: IEEE Photonics Journal.

\bibitem{gilchrist_distance_2005-1}
Alexei Gilchrist, Nathan~K. Langford, and Michael~A. Nielsen.
\newblock Distance measures to compare real and ideal quantum processes.
\newblock {\em Physical Review A}, 71(6):062310, June 2005.
\newblock arXiv:quant-ph/0408063.

\bibitem{stollenwerk_diagrammatic_2022}
Tobias Stollenwerk and Stuart Hadfield.
\newblock Diagrammatic {Analysis} for {Parameterized} {Quantum} {Circuits}.
\newblock Technical Report arXiv:2204.01307, arXiv, April 2022.
\newblock arXiv:2204.01307 [quant-ph] type: article.

\bibitem{jordan_wigner}
P.~Jordan and E.~Wigner.
\newblock Uber das paulische aquivalenzverbot.
\newblock {\em Zeitschrift fur Physik}, 47(9):631--651, 1928.

\bibitem{qasm_bench}
Ang Li, Samuel Stein, Sriram Krishnamoorthy, and James Ang.
\newblock Qasmbench: A low-level quantum benchmark suite for nisq evaluation
  and simulation.
\newblock {\em ACM Transactions on Quantum Computing}, 2022.

\bibitem{stein_2021}
Samuel~A. Stein, Betis Baheri, Daniel Chen, Ying Mao, Qiang Guan, Ang Li,
  Bo~Fang, and Shuai Xu.
\newblock Qugan: A quantum state fidelity based generative adversarial network.
\newblock In {\em 2021 IEEE International Conference on Quantum Computing and
  Engineering (QCE)}, page 71–81. IEEE, October 2021.

\bibitem{draper}
Thomas~G. Draper.
\newblock Addition on a quantum computer, 2000.

\bibitem{bouland_complexity_2019}
Adam Bouland, Bill Fefferman, Chinmay Nirkhe, and Umesh Vazirani.
\newblock On the complexity and verification of quantum random circuit
  sampling.
\newblock {\em Nature Physics}, 15(2):159--163, February 2019.
\newblock Number: 2 Publisher: Nature Publishing Group.

\bibitem{noauthor_new_nodate}
New {Directions} {In} {Testing} {\textbar} {Semantic} {Scholar}.

\bibitem{boixo_characterizing_2018}
Sergio Boixo, Sergei~V. Isakov, Vadim~N. Smelyanskiy, Ryan Babbush, Nan Ding,
  Zhang Jiang, Michael~J. Bremner, John~M. Martinis, and Hartmut Neven.
\newblock Characterizing quantum supremacy in near-term devices.
\newblock {\em Nature Physics}, 14(6):595--600, June 2018.
\newblock Number: 6 Publisher: Nature Publishing Group.

\bibitem{pan_simulating_2021}
Feng Pan and Pan Zhang.
\newblock Simulating the {Sycamore} quantum supremacy circuits, March 2021.
\newblock arXiv:2103.03074 [physics, physics:quant-ph].

\bibitem{arute_supplementary_2019}
Frank Arute, Kunal Arya, Ryan Babbush, Dave Bacon, Joseph~C. Bardin, Rami
  Barends, Rupak Biswas, Sergio Boixo, Fernando G. S.~L. Brandao, David~A.
  Buell, Brian Burkett, Yu~Chen, Zijun Chen, Ben Chiaro, Roberto Collins,
  William Courtney, Andrew Dunsworth, Edward Farhi, Brooks Foxen, Austin
  Fowler, Craig Gidney, Marissa Giustina, Rob Graff, Keith Guerin, Steve
  Habegger, Matthew~P. Harrigan, Michael~J. Hartmann, Alan Ho, Markus~R.
  Hoffmann, Trent Huang, Travis~S. Humble, Sergei~V. Isakov, Evan Jeffrey,
  Zhang Jiang, Dvir Kafri, Kostyantyn Kechedzhi, Julian Kelly, Paul~V. Klimov,
  Sergey Knysh, Alexander~N. Korotkov, Fedor Kostritsa, David Landhuis, Mike
  Lindmark, Erik Lucero, Dmitry Lyakh, Salvatore Mandra, Jarrod~R. McClean,
  Matt McEwen, Anthony Megrant, Xiao Mi, Kristel Michielsen, Masoud Mohseni,
  Josh Mutus, Ofer Naaman, Matthew Neeley, Charles Neill, Murphy~Yuezhen Niu,
  Eric Ostby, Andre Petukhov, John~C. Platt, Chris Quintana, Eleanor~G.
  Rieffel, Pedram Roushan, Nicholas~C. Rubin, Daniel Sank, Kevin~J. Satzinger,
  Vadim Smelyanskiy, Kevin~J. Sung, Matthew~D. Trevithick, Amit Vainsencher,
  Benjamin Villalonga, Theodore White, Z.~Jamie Yao, Ping Yeh, Adam Zalcman,
  Hartmut Neven, and John~M. Martinis.
\newblock Supplementary information for "{Quantum} supremacy using a
  programmable superconducting processor".
\newblock {\em Nature}, 574(7779):505--510, October 2019.
\newblock arXiv:1910.11333 [quant-ph].

\bibitem{caleffi_distributed_2022}
Marcello Caleffi, Michele Amoretti, Davide Ferrari, Daniele Cuomo, Jessica
  Illiano, Antonio Manzalini, and Angela~Sara Cacciapuoti.
\newblock Distributed {Quantum} {Computing}: a {Survey}, December 2022.
\newblock arXiv:2212.10609 [quant-ph].

\bibitem{gidney_how_2021}
Craig Gidney and Martin Ekerå.
\newblock How to factor 2048 bit {RSA} integers in 8 hours using 20 million
  noisy qubits.
\newblock {\em Quantum}, 5:433, April 2021.
\newblock arXiv:1905.09749 [quant-ph].

\bibitem{beauregard_circuit_2003}
Stephane Beauregard.
\newblock Circuit for {Shor}'s algorithm using 2n+3 qubits, February 2003.
\newblock arXiv:quant-ph/0205095.

\bibitem{bravyi_universal_2005}
Sergei Bravyi and Alexei Kitaev.
\newblock Universal {Quantum} {Computation} with ideal {Clifford} gates and
  noisy ancillas.
\newblock {\em Physical Review A}, 71(2):022316, February 2005.
\newblock arXiv:quant-ph/0403025.

\bibitem{herman_survey_2022}
Dylan Herman, Cody Googin, Xiaoyuan Liu, Alexey Galda, Ilya Safro, Yue Sun,
  Marco Pistoia, and Yuri Alexeev.
\newblock A {Survey} of {Quantum} {Computing} for {Finance}, June 2022.
\newblock arXiv:2201.02773 [quant-ph, q-fin].

\bibitem{lee_hybrid_2019}
Yonghae Lee, Jaewoo Joo, and Soojoon Lee.
\newblock Hybrid quantum linear equation algorithm and its experimental test on
  {IBM} {Quantum} {Experience}.
\newblock {\em Scientific Reports}, 9(1):4778, March 2019.
\newblock Number: 1 Publisher: Nature Publishing Group.

\bibitem{gidney_how_2021-1}
Craig Gidney and Martin Ekerå.
\newblock How to factor 2048 bit {RSA} integers in 8 hours using 20 million
  noisy qubits.
\newblock {\em Quantum}, 5:433, April 2021.
\newblock Publisher: Verein zur Förderung des Open Access Publizierens in den
  Quantenwissenschaften.

\bibitem{peruzzo_variational_2014}
Alberto Peruzzo, Jarrod McClean, Peter Shadbolt, Man-Hong Yung, Xiao-Qi Zhou,
  Peter~J. Love, Alán Aspuru-Guzik, and Jeremy~L. O'Brien.
\newblock A variational eigenvalue solver on a quantum processor.
\newblock {\em Nature Communications}, 5(1):4213, July 2014.
\newblock arXiv:1304.3061 [physics, physics:quant-ph].

\bibitem{van_meter_architecture-dependent_2006}
Rodney Van~Meter, Kohei~M. Itoh, and Thaddeus~D. Ladd.
\newblock Architecture-{Dependent} {Execution} {Time} of {Shor}'s {Algorithm},
  May 2006.
\newblock arXiv:quant-ph/0507023.

\bibitem{biamonte_quantum_2017}
Jacob Biamonte, Peter Wittek, Nicola Pancotti, Patrick Rebentrost, Nathan
  Wiebe, and Seth Lloyd.
\newblock Quantum {Machine} {Learning}.
\newblock {\em Nature}, 549(7671):195--202, September 2017.
\newblock arXiv:1611.09347 [cond-mat, physics:quant-ph, stat].

\bibitem{harrow_quantum_2009}
Aram~W. Harrow, Avinatan Hassidim, and Seth Lloyd.
\newblock Quantum algorithm for solving linear systems of equations.
\newblock {\em Physical Review Letters}, 103(15):150502, October 2009.
\newblock arXiv:0811.3171 [quant-ph].

\bibitem{van_leent_entangling_2022}
Tim van Leent, Matthias Bock, Florian Fertig, Robert Garthoff, Sebastian
  Eppelt, Yiru Zhou, Pooja Malik, Matthias Seubert, Tobias Bauer, Wenjamin
  Rosenfeld, Wei Zhang, Christoph Becher, and Harald Weinfurter.
\newblock Entangling single atoms over 33 km telecom fibre.
\newblock {\em Nature}, 607(7917):69--73, July 2022.
\newblock Number: 7917 Publisher: Nature Publishing Group.

\bibitem{van_leent_entangling_2022-1}
Tim van Leent, Matthias Bock, Florian Fertig, Robert Garthoff, Sebastian
  Eppelt, Yiru Zhou, Pooja Malik, Matthias Seubert, Tobias Bauer, Wenjamin
  Rosenfeld, Wei Zhang, Christoph Becher, and Harald Weinfurter.
\newblock Entangling single atoms over 33 km telecom fibre.
\newblock {\em Nature}, 607(7917):69--73, July 2022.

\bibitem{cleve_how_1999}
Richard Cleve, Daniel Gottesman, and Hoi-Kwong Lo.
\newblock How to share a quantum secret.
\newblock {\em Physical Review Letters}, 83(3):648--651, July 1999.
\newblock arXiv:quant-ph/9901025.

\bibitem{haner_distributed_2021}
Thomas Häner, Damian~S. Steiger, Torsten Hoefler, and Matthias Troyer.
\newblock Distributed {Quantum} {Computing} with {QMPI}.
\newblock In {\em Proceedings of the {International} {Conference} for {High}
  {Performance} {Computing}, {Networking}, {Storage} and {Analysis}}, pages
  1--13, November 2021.
\newblock arXiv:2105.01109 [quant-ph].

\bibitem{zhang_minimum_2004}
Jun Zhang, Jiri Vala, Shankar Sastry, and K.~Birgitta Whaley.
\newblock Minimum construction of two-qubit quantum operations.
\newblock {\em Physical Review Letters}, 93(2):020502, July 2004.
\newblock arXiv:quant-ph/0312193.

\bibitem{kim_evidence_2023}
Youngseok Kim, Andrew Eddins, Sajant Anand, Ken~Xuan Wei, Ewout van~den Berg,
  Sami Rosenblatt, Hasan Nayfeh, Yantao Wu, Michael Zaletel, Kristan Temme, and
  Abhinav Kandala.
\newblock Evidence for the utility of quantum computing before fault tolerance.
\newblock {\em Nature}, 618(7965):500--505, June 2023.
\newblock Number: 7965 Publisher: Nature Publishing Group.

\bibitem{bao_fluxonium_2022}
Feng Bao, Hao Deng, Dawei Ding, Ran Gao, Xun Gao, Cupjin Huang, Xun Jiang,
  Hsiang-Sheng Ku, Zhisheng Li, Xizheng Ma, Xiaotong Ni, Jin Qin, Zhijun Song,
  Hantao Sun, Chengchun Tang, Tenghui Wang, Feng Wu, Tian Xia, Wenlong Yu, Fang
  Zhang, Gengyan Zhang, Xiaohang Zhang, Jingwei Zhou, Xing Zhu, Yaoyun Shi,
  Jianxin Chen, Hui-Hai Zhao, and Chunqing Deng.
\newblock Fluxonium: an alternative qubit platform for high-fidelity
  operations.
\newblock {\em Physical Review Letters}, 129(1):010502, June 2022.
\newblock arXiv:2111.13504 [quant-ph].

\bibitem{chen_benchmarking_2023}
Jwo-Sy Chen, Erik Nielsen, Matthew Ebert, Volkan Inlek, Kenneth Wright,
  Vandiver Chaplin, Andrii Maksymov, Eduardo Páez, Amrit Poudel, Peter Maunz,
  and John Gamble.
\newblock Benchmarking a trapped-ion quantum computer with 29 algorithmic
  qubits, August 2023.
\newblock arXiv:2308.05071 [quant-ph].

\bibitem{zhang_minimum_2004-1}
Jun Zhang, Jiri Vala, Shankar Sastry, and K.~Birgitta Whaley.
\newblock Minimum construction of two-qubit quantum operations.
\newblock {\em Physical Review Letters}, 93(2):020502, July 2004.
\newblock arXiv:quant-ph/0312193.

\bibitem{akhtar_high-fidelity_2023}
M.~Akhtar, F.~Bonus, F.~R. Lebrun-Gallagher, N.~I. Johnson, M.~Siegele-Brown,
  S.~Hong, S.~J. Hile, S.~A. Kulmiya, S.~Weidt, and W.~K. Hensinger.
\newblock A high-fidelity quantum matter-link between ion-trap microchip
  modules.
\newblock {\em Nature Communications}, 14(1):531, February 2023.
\newblock Number: 1 Publisher: Nature Publishing Group.

\bibitem{magesan_robust_2011}
Easwar Magesan, J.~M. Gambetta, and Joseph Emerson.
\newblock Robust randomized benchmarking of quantum processes.
\newblock {\em Physical Review Letters}, 106(18):180504, May 2011.
\newblock arXiv:1009.3639 [quant-ph].

\bibitem{magesan_characterizing_2012}
Easwar Magesan, Jay~M. Gambetta, and Joseph Emerson.
\newblock Characterizing {Quantum} {Gates} via {Randomized} {Benchmarking}.
\newblock {\em Physical Review A}, 85(4):042311, April 2012.
\newblock arXiv:1109.6887 [quant-ph].

\bibitem{knill_randomized_2008}
E.~Knill, D.~Leibfried, R.~Reichle, J.~Britton, R.~B. Blakestad, J.~D. Jost,
  C.~Langer, R.~Ozeri, S.~Seidelin, and D.~J. Wineland.
\newblock Randomized {Benchmarking} of {Quantum} {Gates}.
\newblock {\em Physical Review A}, 77(1):012307, January 2008.
\newblock arXiv:0707.0963 [quant-ph].

\bibitem{arute_supplementary_2019-1}
Frank Arute, Kunal Arya, Ryan Babbush, Dave Bacon, Joseph~C. Bardin, Rami
  Barends, Rupak Biswas, Sergio Boixo, Fernando G. S.~L. Brandao, David~A.
  Buell, Brian Burkett, Yu~Chen, Zijun Chen, Ben Chiaro, Roberto Collins,
  William Courtney, Andrew Dunsworth, Edward Farhi, Brooks Foxen, Austin
  Fowler, Craig Gidney, Marissa Giustina, Rob Graff, Keith Guerin, Steve
  Habegger, Matthew~P. Harrigan, Michael~J. Hartmann, Alan Ho, Markus~R.
  Hoffmann, Trent Huang, Travis~S. Humble, Sergei~V. Isakov, Evan Jeffrey,
  Zhang Jiang, Dvir Kafri, Kostyantyn Kechedzhi, Julian Kelly, Paul~V. Klimov,
  Sergey Knysh, Alexander~N. Korotkov, Fedor Kostritsa, David Landhuis, Mike
  Lindmark, Erik Lucero, Dmitry Lyakh, Salvatore Mandra, Jarrod~R. McClean,
  Matt McEwen, Anthony Megrant, Xiao Mi, Kristel Michielsen, Masoud Mohseni,
  Josh Mutus, Ofer Naaman, Matthew Neeley, Charles Neill, Murphy~Yuezhen Niu,
  Eric Ostby, Andre Petukhov, John~C. Platt, Chris Quintana, Eleanor~G.
  Rieffel, Pedram Roushan, Nicholas~C. Rubin, Daniel Sank, Kevin~J. Satzinger,
  Vadim Smelyanskiy, Kevin~J. Sung, Matthew~D. Trevithick, Amit Vainsencher,
  Benjamin Villalonga, Theodore White, Z.~Jamie Yao, Ping Yeh, Adam Zalcman,
  Hartmut Neven, and John~M. Martinis.
\newblock Supplementary information for "{Quantum} supremacy using a
  programmable superconducting processor".
\newblock {\em Nature}, 574(7779):505--510, October 2019.
\newblock arXiv:1910.11333 [quant-ph].

\bibitem{cross_validating_2019}
Andrew~W. Cross, Lev~S. Bishop, Sarah Sheldon, Paul~D. Nation, and Jay~M.
  Gambetta.
\newblock Validating quantum computers using randomized model circuits.
\newblock {\em Physical Review A}, 100(3):032328, September 2019.
\newblock arXiv:1811.12926 [quant-ph].

\bibitem{younis_quantum_2022}
Ed~Younis and Costin Iancu.
\newblock Quantum {Circuit} {Optimization} and {Transpilation} via
  {Parameterized} {Circuit} {Instantiation}, June 2022.
\newblock arXiv:2206.07885 [quant-ph].

\bibitem{tket}
Seyon Sivarajah, Silas Dilkes, Alexander Cowtan, Will Simmons, Alec Edgington,
  and Ross Duncan.
\newblock tket : {A} {Retargetable} {Compiler} for {NISQ} {Devices}.
\newblock {\em Quantum Science and Technology}, 6(1):014003, January 2021.
\newblock arXiv:2003.10611 [quant-ph].

\bibitem{cirq}
Cirq Developers.
\newblock Cirq, July 2023.

\bibitem{wille_mqt_2023}
Robert Wille and Lukas Burgholzer.
\newblock {MQT} {QMAP}: {Efficient} {Quantum} {Circuit} {Mapping}.
\newblock In {\em Proceedings of the 2023 {International} {Symposium} on
  {Physical} {Design}}, pages 198--204, March 2023.
\newblock arXiv:2301.11935 [quant-ph].

\bibitem{horodecki_general_1999}
Pawel Horodecki, Michal Horodecki, and Ryszard Horodecki.
\newblock General teleportation channel, singlet fraction and
  quasi-distillation, March 1999.
\newblock arXiv:quant-ph/9807091.

\bibitem{weiden_improving_2023}
Mathias Weiden, Ed~Younis, Justin Kalloor, John Kubiatowicz, and Costin Iancu.
\newblock Improving {Quantum} {Circuit} {Synthesis} with {Machine} {Learning},
  June 2023.
\newblock arXiv:2306.05622 [quant-ph].

\bibitem{ross_sellinger}
Neil~J. Ross and Peter Selinger.
\newblock Optimal ancilla-free clifford+t approximation of z-rotations, 2016.

\bibitem{mandviwalla_implementing_2018}
Aamir Mandviwalla, Keita Ohshiro, and Bo~Ji.
\newblock Implementing {Grover}’s {Algorithm} on the {IBM} {Quantum}
  {Computers}.
\newblock In {\em 2018 {IEEE} {International} {Conference} on {Big} {Data}
  ({Big} {Data})}, pages 2531--2537, December 2018.

\bibitem{farhi_quantum_2014}
Edward Farhi, Jeffrey Goldstone, and Sam Gutmann.
\newblock A {Quantum} {Approximate} {Optimization} {Algorithm}, November 2014.
\newblock arXiv:1411.4028 [quant-ph].

\bibitem{bravyi_fermionic_2002}
Sergey Bravyi and Alexei Kitaev.
\newblock Fermionic quantum computation.
\newblock {\em Annals of Physics}, 298(1):210--226, May 2002.
\newblock arXiv:quant-ph/0003137.

\bibitem{peruzzo_variational_2014-1}
Alberto Peruzzo, Jarrod McClean, Peter Shadbolt, Man-Hong Yung, Xiao-Qi Zhou,
  Peter~J. Love, Alán Aspuru-Guzik, and Jeremy~L. O'Brien.
\newblock A variational eigenvalue solver on a quantum processor.
\newblock {\em Nature Communications}, 5(1):4213, July 2014.
\newblock arXiv:1304.3061 [physics, physics:quant-ph].

\bibitem{shin_phonon-driven_2018}
Dongbin Shin, Hannes Hübener, Umberto De~Giovannini, Hosub Jin, Angel Rubio,
  and Noejung Park.
\newblock Phonon-driven spin-{Floquet} magneto-valleytronics in {MoS2}.
\newblock {\em Nature Communications}, 9(1):638, February 2018.
\newblock Number: 1 Publisher: Nature Publishing Group.

\bibitem{li_co-design_2021}
Gushu Li, Anbang Wu, Yunong Shi, Ali Javadi-Abhari, Yufei Ding, and Yuan Xie.
\newblock On the {Co}-{Design} of {Quantum} {Software} and {Hardware}.
\newblock In {\em Proceedings of the {Eight} {Annual} {ACM} {International}
  {Conference} on {Nanoscale} {Computing} and {Communication}}, {NANOCOM} '21,
  pages 1--7, New York, NY, USA, September 2021. Association for Computing
  Machinery.

\bibitem{safi_influence_2023}
Hila Safi, Karen Wintersperger, and Wolfgang Mauerer.
\newblock Influence of {HW}-{SW}-{Co}-{Design} on {Quantum} {Computing}
  {Scalability}, June 2023.
\newblock arXiv:2306.04246 [quant-ph].

\bibitem{erhard_characterizing_2019}
Alexander Erhard, Joel~James Wallman, Lukas Postler, Michael Meth, Roman
  Stricker, Esteban~Adrian Martinez, Philipp Schindler, Thomas Monz, Joseph
  Emerson, and Rainer Blatt.
\newblock Characterizing large-scale quantum computers via cycle benchmarking.
\newblock {\em Nature Communications}, 10(1):5347, November 2019.
\newblock arXiv:1902.08543 [quant-ph].

\bibitem{cerezo_challenges_2022}
M.~Cerezo, Guillaume Verdon, Hsin-Yuan Huang, Lukasz Cincio, and Patrick~J.
  Coles.
\newblock Challenges and opportunities in quantum machine learning.
\newblock {\em Nature Computational Science}, 2(9):567--576, September 2022.
\newblock Number: 9 Publisher: Nature Publishing Group.

\bibitem{noauthor_ionq_nodate}
{IonQ} {Forte}: {The} {First} {Software}-{Configurable} {Quantum} {Computer}.

\bibitem{noauthor_rigetti_nodate}
Rigetti {QCS}.

\bibitem{noauthor_quantinuum_nodate}
Quantinuum {\textbar} {Hardware} {\textbar} {System} {Model} {H2}.

\bibitem{noauthor_technology_nodate}
Technology.

\bibitem{stein_qugan_2021}
Samuel~A. Stein, Betis Baheri, Daniel Chen, Ying Mao, Qiang Guan, Ang Li,
  Bo~Fang, and Shuai Xu.
\newblock {QuGAN}: {A} {Quantum} {State} {Fidelity} based {Generative}
  {Adversarial} {Network}.
\newblock In {\em 2021 {IEEE} {International} {Conference} on {Quantum}
  {Computing} and {Engineering} ({QCE})}, pages 71--81, October 2021.
\newblock arXiv:2010.09036 [quant-ph].

\bibitem{hartree_fock}
Attila Szabo and Neil~S. Ostlund.
\newblock {\em Modern Quantum Chemistry: Introduction to Advanced Electronic
  Structure Theory}.
\newblock Dover Publications, Inc., Mineola, first edition, 1996.

\bibitem{brassard}
Gilles Brassard, Peter Hoyer, Michele Mosca, and Alain Tapp.
\newblock Quantum {Amplitude} {Amplification} and {Estimation}.
\newblock volume 305, pages 53--74. 2002.
\newblock arXiv:quant-ph/0005055.

\bibitem{rahman_2022}
Sarmed~A. Rahman, Randy Lewis, Emanuele Mendicelli, and Sarah Powell.
\newblock Self-mitigating {Trotter} circuits for {SU}(2) lattice gauge theory
  on a quantum computer.
\newblock {\em Physical Review D}, 106(7):074502, October 2022.
\newblock arXiv:2205.09247 [hep-lat].

\bibitem{brassard_quantum_2002-1}
Gilles Brassard, Peter Hoyer, Michele Mosca, and Alain Tapp.
\newblock Quantum {Amplitude} {Amplification} and {Estimation}.
\newblock volume 305, pages 53--74. 2002.
\newblock arXiv:quant-ph/0005055.

\end{thebibliography}


\begin{thebibliography}{10}
\urlstyle{rm}
\expandafter\ifx\csname url\endcsname\relax
  \def\url#1{\texttt{#1}}\fi
\expandafter\ifx\csname urlprefix\endcsname\relax\def\urlprefix{URL }\fi
\expandafter\ifx\csname doiprefix\endcsname\relax\def\doiprefix{DOI: }\fi
\providecommand{\bibinfo}[2]{#2}
\providecommand{\eprint}[2][]{\url{#2}}

\bibitem{Preskill2018_NISQ}
\bibinfo{author}{Preskill, J.}
\newblock \bibinfo{journal}{\bibinfo{title}{Quantum {C}omputing in the {NISQ}
  era and beyond}}.
\newblock {\emph{\JournalTitle{{Quantum}}}} \textbf{\bibinfo{volume}{2}},
  \bibinfo{pages}{79}, \doiprefix\url{10.22331/q-2018-08-06-79}
  (\bibinfo{year}{2018}).

\bibitem{botea2018complexity}
\bibinfo{author}{Botea, A.}, \bibinfo{author}{Kishimoto, A.} \&
  \bibinfo{author}{Marinescu, R.}
\newblock \bibinfo{title}{On the complexity of quantum circuit compilation}.
\newblock In \emph{\bibinfo{booktitle}{Proceedings of the International
  Symposium on Combinatorial Search}}, vol.~\bibinfo{volume}{9},
  \bibinfo{pages}{138--142} (\bibinfo{year}{2018}).

\bibitem{qiskit}
\bibinfo{author}{Javadi-Abhari, A.} \emph{et~al.}
\newblock \bibinfo{title}{Quantum computing with {Q}iskit},
  \doiprefix\url{10.48550/arXiv.2405.08810} (\bibinfo{year}{2024}).
\newblock \eprint{2405.08810}.

\bibitem{bqskit}
\bibinfo{author}{Younis, E.}, \bibinfo{author}{Iancu, C.~C.},
  \bibinfo{author}{Lavrijsen, W.}, \bibinfo{author}{Davis, M.} \&
  \bibinfo{author}{Smith, E.}
\newblock \bibinfo{title}{Berkeley quantum synthesis toolkit (bqskit) v1}.
\newblock \bibinfo{howpublished}{[Computer Software]
  \url{https://doi.org/10.11578/dc.20210603.2}},
  \doiprefix\url{10.11578/dc.20210603.2} (\bibinfo{year}{2021}).

\bibitem{tket}
\bibinfo{author}{Sivarajah, S.} \emph{et~al.}
\newblock \bibinfo{journal}{\bibinfo{title}{tket : {A} {Retargetable}
  {Compiler} for {NISQ} {Devices}}}.
\newblock {\emph{\JournalTitle{Quantum Science and Technology}}}
  \textbf{\bibinfo{volume}{6}}, \bibinfo{pages}{014003},
  \doiprefix\url{10.1088/2058-9565/ab8e92} (\bibinfo{year}{2021}).
\newblock \bibinfo{note}{ArXiv:2003.10611 [quant-ph]}.

\bibitem{cirq}
\bibinfo{author}{Developers, C.}
\newblock \bibinfo{title}{Cirq}, \doiprefix\url{10.5281/zenodo.8161252}
  (\bibinfo{year}{2023}).

\bibitem{qdrift}
\bibinfo{author}{Campbell, E.}
\newblock \bibinfo{journal}{\bibinfo{title}{Random compiler for fast
  hamiltonian simulation}}.
\newblock {\emph{\JournalTitle{Physical Review Letters}}}
  \textbf{\bibinfo{volume}{123}},
  \doiprefix\url{10.1103/physrevlett.123.070503} (\bibinfo{year}{2019}).

\bibitem{PRXQuantum.5.020101}
\bibinfo{author}{Katabarwa, A.}, \bibinfo{author}{Gratsea, K.},
  \bibinfo{author}{Caesura, A.} \& \bibinfo{author}{Johnson, P.~D.}
\newblock \bibinfo{journal}{\bibinfo{title}{Early fault-tolerant quantum
  computing}}.
\newblock {\emph{\JournalTitle{PRX Quantum}}} \textbf{\bibinfo{volume}{5}},
  \bibinfo{pages}{020101}, \doiprefix\url{10.1103/PRXQuantum.5.020101}
  (\bibinfo{year}{2024}).

\bibitem{wu_qgo}
\bibinfo{author}{Wu, X.-C.}, \bibinfo{author}{Davis, M.~G.},
  \bibinfo{author}{Chong, F.~T.} \& \bibinfo{author}{Iancu, C.}
\newblock \bibinfo{title}{Reoptimization of quantum circuits via hierarchical
  synthesis}.
\newblock In \emph{\bibinfo{booktitle}{2021 International Conference on
  Rebooting Computing (ICRC)}}, \bibinfo{pages}{35--46},
  \doiprefix\url{10.1109/ICRC53822.2021.00016} (\bibinfo{year}{2021}).

\bibitem{Burt_2024}
\bibinfo{author}{Burt, F.}, \bibinfo{author}{Chen, K.-C.} \&
  \bibinfo{author}{Leung, K.~K.}
\newblock \bibinfo{title}{Generalised circuit partitioning for distributed
  quantum computing}.
\newblock In \emph{\bibinfo{booktitle}{2024 IEEE International Conference on
  Quantum Computing and Engineering (QCE)}}, \bibinfo{pages}{173–178},
  \doiprefix\url{10.1109/qce60285.2024.10273} (\bibinfo{publisher}{IEEE},
  \bibinfo{year}{2024}).

\bibitem{PhysRevA.95.042306}
\bibinfo{author}{Campbell, E.}
\newblock \bibinfo{journal}{\bibinfo{title}{Shorter gate sequences for quantum
  computing by mixing unitaries}}.
\newblock {\emph{\JournalTitle{Phys. Rev. A}}} \textbf{\bibinfo{volume}{95}},
  \bibinfo{pages}{042306}, \doiprefix\url{10.1103/PhysRevA.95.042306}
  (\bibinfo{year}{2017}).

\bibitem{hastings2016turning}
\bibinfo{author}{Hastings, M.~B.}
\newblock \bibinfo{journal}{\bibinfo{title}{Turning gate synthesis errors into
  incoherent errors}}.
\newblock {\emph{\JournalTitle{arXiv:1612.01011}}}  (\bibinfo{year}{2016}).

\bibitem{Kliuchnikov_2023}
\bibinfo{author}{Kliuchnikov, V.}, \bibinfo{author}{Lauter, K.},
  \bibinfo{author}{Minko, R.}, \bibinfo{author}{Paetznick, A.} \&
  \bibinfo{author}{Petit, C.}
\newblock \bibinfo{journal}{\bibinfo{title}{Shorter quantum circuits via
  single-qubit gate approximation}}.
\newblock {\emph{\JournalTitle{{Quantum}}}} \textbf{\bibinfo{volume}{7}},
  \bibinfo{pages}{1208}, \doiprefix\url{10.22331/q-2023-12-18-1208}
  (\bibinfo{year}{2023}).

\bibitem{Akibu_ACM_2024}
\bibinfo{author}{Akibue, S.}, \bibinfo{author}{Kato, G.} \&
  \bibinfo{author}{Tani, S.}
\newblock \bibinfo{journal}{\bibinfo{title}{Probabilistic unitary synthesis
  with optimal accuracy}}.
\newblock {\emph{\JournalTitle{ACM Transactions on Quantum Computing}}}
  \textbf{\bibinfo{volume}{5}}, \doiprefix\url{10.1145/3663576}
  (\bibinfo{year}{2024}).

\bibitem{yoshioka2024errorcrafting}
\bibinfo{author}{Yoshioka, N.}, \bibinfo{author}{Akibue, S.},
  \bibinfo{author}{Morisaki, H.}, \bibinfo{author}{Tsubouchi, K.} \&
  \bibinfo{author}{Suzuki, Y.}
\newblock \bibinfo{title}{Error crafting in probabilistic quantum gate
  synthesis} (\bibinfo{year}{2024}).
\newblock \eprint{2405.15565}.

\bibitem{Akibue2024}
\bibinfo{author}{Akibue, S.}, \bibinfo{author}{Kato, G.} \&
  \bibinfo{author}{Tani, S.}
\newblock \bibinfo{journal}{\bibinfo{title}{Probabilistic state synthesis based
  on optimal convex approximation}}.
\newblock {\emph{\JournalTitle{npj Quantum Information}}}
  \textbf{\bibinfo{volume}{10}}, \bibinfo{pages}{3},
  \doiprefix\url{10.1038/s41534-023-00793-7} (\bibinfo{year}{2024}).

\bibitem{low2021halving}
\bibinfo{author}{Low, G.~H.}
\newblock \bibinfo{title}{Halving the cost of quantum multiplexed rotations}
  (\bibinfo{year}{2021}).
\newblock \eprint{2110.13439}.

\bibitem{campbell_qdrift}
\bibinfo{author}{Campbell, E.}
\newblock \bibinfo{journal}{\bibinfo{title}{Random compiler for fast
  hamiltonian simulation}}.
\newblock {\emph{\JournalTitle{Phys. Rev. Lett.}}}
  \textbf{\bibinfo{volume}{123}}, \bibinfo{pages}{070503},
  \doiprefix\url{10.1103/PhysRevLett.123.070503} (\bibinfo{year}{2019}).

\bibitem{Ouyang2020compilation}
\bibinfo{author}{Ouyang, Y.}, \bibinfo{author}{White, D.~R.} \&
  \bibinfo{author}{Campbell, E.~T.}
\newblock \bibinfo{journal}{\bibinfo{title}{Compilation by stochastic
  {H}amiltonian sparsification}}.
\newblock {\emph{\JournalTitle{{Quantum}}}} \textbf{\bibinfo{volume}{4}},
  \bibinfo{pages}{235}, \doiprefix\url{10.22331/q-2020-02-27-235}
  (\bibinfo{year}{2020}).

\bibitem{PRXQuantum.2.040305}
\bibinfo{author}{Chen, C.-F.}, \bibinfo{author}{Huang, H.-Y.},
  \bibinfo{author}{Kueng, R.} \& \bibinfo{author}{Tropp, J.~A.}
\newblock \bibinfo{journal}{\bibinfo{title}{Concentration for random product
  formulas}}.
\newblock {\emph{\JournalTitle{PRX Quantum}}} \textbf{\bibinfo{volume}{2}},
  \bibinfo{pages}{040305}, \doiprefix\url{10.1103/PRXQuantum.2.040305}
  (\bibinfo{year}{2021}).

\bibitem{huang_2023}
\bibinfo{author}{Huang, Y.}, \bibinfo{author}{Shao, Y.}, \bibinfo{author}{Ren,
  W.}, \bibinfo{author}{Sun, J.} \& \bibinfo{author}{Lv, D.}
\newblock \bibinfo{journal}{\bibinfo{title}{Efficient quantum imaginary time
  evolution by drifting real-time evolution: An approach with low gate and
  measurement complexity}}.
\newblock {\emph{\JournalTitle{Journal of Chemical Theory and Computation}}}
  \textbf{\bibinfo{volume}{19}}, \bibinfo{pages}{3868--3876},
  \doiprefix\url{10.1021/acs.jctc.3c00071} (\bibinfo{year}{2023}).
\newblock \eprint{https://doi.org/10.1021/acs.jctc.3c00071}.

\bibitem{PhysRevResearch.6.013224}
\bibinfo{author}{Pocrnic, M.}, \bibinfo{author}{Hagan, M.},
  \bibinfo{author}{Carrasquilla, J.}, \bibinfo{author}{Segal, D.} \&
  \bibinfo{author}{Wiebe, N.}
\newblock \bibinfo{journal}{\bibinfo{title}{Composite qdrift-product formulas
  for quantum and classical simulations in real and imaginary time}}.
\newblock {\emph{\JournalTitle{Phys. Rev. Res.}}} \textbf{\bibinfo{volume}{6}},
  \bibinfo{pages}{013224}, \doiprefix\url{10.1103/PhysRevResearch.6.013224}
  (\bibinfo{year}{2024}).

\bibitem{PhysRevA.109.062431}
\bibinfo{author}{Cho, C.-H.}, \bibinfo{author}{Berry, D.~W.} \&
  \bibinfo{author}{Hsieh, M.-H.}
\newblock \bibinfo{journal}{\bibinfo{title}{Doubling the order of approximation
  via the randomized product formula}}.
\newblock {\emph{\JournalTitle{Phys. Rev. A}}} \textbf{\bibinfo{volume}{109}},
  \bibinfo{pages}{062431}, \doiprefix\url{10.1103/PhysRevA.109.062431}
  (\bibinfo{year}{2024}).

\bibitem{martyn2024}
\bibinfo{author}{Martyn, J.~M.} \& \bibinfo{author}{Rall, P.}
\newblock \bibinfo{title}{Halving the cost of quantum algorithms with
  randomization} (\bibinfo{year}{2024}).
\newblock \eprint{2409.03744}.

\bibitem{patel_2021}
\bibinfo{author}{Patel, T.}, \bibinfo{author}{Younis, E.},
  \bibinfo{author}{Iancu, C.}, \bibinfo{author}{de~Jong, W.} \&
  \bibinfo{author}{Tiwari, D.}
\newblock \bibinfo{title}{Robust and resource-efficient quantum circuit
  approximation} (\bibinfo{year}{2021}).
\newblock \eprint{2108.12714}.

\bibitem{watrous_2018}
\bibinfo{author}{Watrous, J.}
\newblock \emph{\bibinfo{title}{The Theory of Quantum Information}}
  (\bibinfo{publisher}{Cambridge University Press}, \bibinfo{year}{2018}).

\bibitem{ross_sellinger}
\bibinfo{author}{Ross, N.~J.} \& \bibinfo{author}{Selinger, P.}
\newblock \bibinfo{title}{Optimal ancilla-free clifford+t approximation of
  z-rotations} (\bibinfo{year}{2016}).
\newblock \eprint{1403.2975}.

\bibitem{Ueda_1996}
\bibinfo{author}{Ueda, N.} \& \bibinfo{author}{Nakano, R.}
\newblock \bibinfo{title}{Generalization error of ensemble estimators}.
\newblock In \emph{\bibinfo{booktitle}{Proceedings of International Conference
  on Neural Networks (ICNN'96)}}, vol.~\bibinfo{volume}{1},
  \bibinfo{pages}{90--95 vol.1}, \doiprefix\url{10.1109/ICNN.1996.548872}
  (\bibinfo{year}{1996}).

\bibitem{Brown_20005}
\bibinfo{author}{Brown, G.}, \bibinfo{author}{Wyatt, J.~L.} \&
  \bibinfo{author}{Ti{\~n}o, P.}
\newblock \bibinfo{journal}{\bibinfo{title}{Managing diversity in regression
  ensembles}}.
\newblock {\emph{\JournalTitle{Journal of Machine Learning Research}}}
  \textbf{\bibinfo{volume}{6}}, \bibinfo{pages}{1621--1650}
  (\bibinfo{year}{2005}).

\bibitem{ntro}
\bibinfo{author}{Davis, M.~G.}
\newblock \emph{\bibinfo{title}{Numerical Synthesis of Arbitrary Multi-Qubit
  Unitaries with low T-Count}}.
\newblock Ph.D. thesis, \bibinfo{school}{Massachusetts Institute of Technology}
  (\bibinfo{year}{2023}).
\newblock \doiprefix\url{10.13140/RG.2.2.32303.30886}.

\bibitem{leap}
\bibinfo{author}{Smith, E.} \emph{et~al.}
\newblock \bibinfo{journal}{\bibinfo{title}{Leap: Scaling numerical
  optimization based synthesis using an incremental approach}}.
\newblock {\emph{\JournalTitle{ACM Transactions on Quantum Computing}}}
  \textbf{\bibinfo{volume}{4}}, \doiprefix\url{10.1145/3548693}
  (\bibinfo{year}{2023}).

\bibitem{qsearch}
\bibinfo{author}{Davis, M.~G.} \emph{et~al.}
\newblock \bibinfo{title}{Towards optimal topology aware quantum circuit
  synthesis}.
\newblock In \emph{\bibinfo{booktitle}{2020 IEEE International Conference on
  Quantum Computing and Engineering (QCE)}}, \bibinfo{pages}{223--234},
  \doiprefix\url{10.1109/QCE49297.2020.00036} (\bibinfo{year}{2020}).

\end{thebibliography}
\end{document}


\flushbottom
\maketitle

\section{Description of benchmark applications and circuits}
For our benchmarks, we use a variety of quantum algorithms across different domains of interest.

\subsection*{Hamiltonian Simulation Algorithms}
\begin{enumerate}
    \item \textbf{SU(2) Lattice Gague Simulation} - We follow the construction outlined in the paper by Rahman et. al \cite{rahman_2022} to simulate the motion of an excitation on a spatial lattice.
    \item \textbf{Li-H Molecule} - We simulate the electronic energy of a Lithium Hydride molecule. The circuit was generated using the Qiskit Nature and PySCF\cite{qiskit} libraries. The fermionic operator was mapped to qubits using a Jordan Wigner mapper\cite{jordan_wigner}. When calculating the observable, we use the Hartree Fock \cite{hartree_fock} state as the initial state.
    \item \textbf{Fermi-Hubbard Model} We run the Fermi-Hubbard model on a square lattice with periodic boundary conditions. The lattice has uniform interaction energies as well as a uniform onsite potential and was mapped to a quantum circuit with a Jordan Wigner mapper. For the observable calculations, we pass in the ground state of the non-interacting model as the initial state.
    \item \textbf{Heisenberg Model} - We simulate the nearest neighbor Heisenberg model with the Hamiltonian defined as:
    \[H = J \cdot \sum{X_iX_{i+1} + Y_iY_{i+1} + Z_iZ_{i+1}}\] and use the all zero state as the initial ground state.
\end{enumerate}

\subsection*{Arithmetic Sub-Modules}

\begin{enumerate}
    \item \textbf{Multiplier} - A quantum sub-circuit to multiply two registers and store into an output register\cite{qiskit}.
    \item \textbf{Adder} - A simple implementation of a Ripple-Carry Adder to add together two quantum registers \cite{qiskit}.
    \item \textbf{QFT Adder} - An implementation of the Draper QFT Adder \cite{draper}.
\end{enumerate}

\subsection*{Quantum Sub-Routines}

\begin{enumerate}
    \item \textbf{Quantum Amplitude Estimation (QAE)} - An implementation of the quantum amplitude estimation circuit on a Bernoulli Operator with a probability of 0.5 \cite{qiskit, brassard}.
    \item \textbf{Quantum Phase Estimation (QPE)} - We estimate the phase of a randomly generated circuit. In our 14 qubit example, we generate a 7-qubit random circuit, and use 7 qubits for precision.
\end{enumerate}

\subsection*{Quantum Optimization}

\begin{enumerate}
    \item Quantum Approximate Optimization Algorithm - An implementation of the optimization circuit given in QASMBench\cite{qasm_bench}.
\end{enumerate}

\section{``Rigid" Circuit Blocks}
\label{sec:rigid_blocks}
As ensemble generation time grows exponentially with the number of qubits, circuit partitioning allows us to scale our workflow to application-width circuits. We first split our circuit into blocks of width $w$, and then generate a block-level ensemble for each partition \emph{in parallel}. Importantly, we are able to accept/reject each ensemble based on the ReWEE condition, which gives us additional flexibility on which circuits benefit from our methodology. We do not need a perfect ensemble for the circuit as a whole; we can pick and choose which parts of the circuit benefit from approximation.

Despite this additional freedom provided by the partitioner, there are benchmarks that are unable to see any reduction in the NISQ compilation workflow. This happens when the input circuit is partitioned into ``rigid" blocks. Rigid blocks are circuits (along with their underlying unitaries) that either 1) Cannot be reduced (e.g no CNOTs can be removed without significant approximation error or 2) Only have a few approximate solutions, which are not diverse enough to form a convex hull around the original unitary.

\begin{figure}[htbp]
\centering
\includegraphics[width=0.55\textwidth]{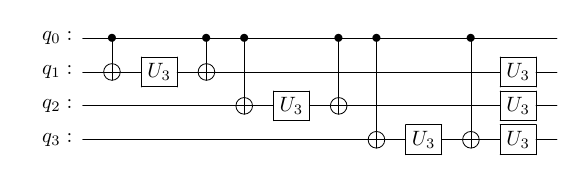} \\
\includegraphics[width=0.8\textwidth]{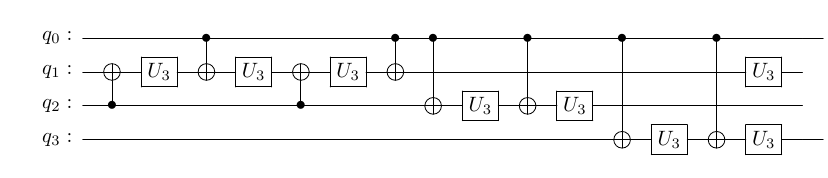}
\caption{Two examples of ``rigid" circuit blocks from the QFT Adder circuit and QPE circuit.}
\label{fig:rigid_blocks}
\end{figure}

Figure \ref{fig:rigid_blocks} illustrates two examples of these rigid structures. The first block is from the QFT Adder circuit while the second block is from a QPE circuit. The first circuit contains 3 consecutive controlled rotations with the same control qubit (q0), while the second circuit contains a multi-controlled rotation and two controlled rotations. Consecutive rotations with a only a few control qubits prove difficult to reduce in the NISQ setting, since the removal of any single CNOT adds significant error to the underlying unitary. On top of that, since the control qubit is being constantly reused, it is hard to find approximate ``shortcuts" by leveraging CNOTs on surrounding qubits.

For the demonstration of our workflow shown in this work, we use a block size of $4$, while the standard BQSKit compile width is $3$. We found that by increasing the block size, both the reduction quality and probability of finding a convex hull improved, as the partitions were more likely to have many control and target qubits and be less rigid. We predict that increasing the block size to larger and larger sizes would continue to improve our results, albeit at the cost of compilation time.

\section{Statistics of ensemble error reduction}
\label{sec:error_scaling}

The benefits of using an ensemble channel are guaranteed only if the ReWEE condition is satisfied, \ie the weighted ensemble error is quadratically reduced, 
\begin{align}
    \normf{\sum_{i=1}^{M^\k} p^\k_i U^\k_i - V^\k} \leq O(\epsilon^2), 
\label{eq:wee_cond}
\end{align}
when each compilation, $U^\k_i$, is performed to error $\epsilon$.  
In practice, it is challenging to construct an ensemble of approximate circuits that both:  
\begin{enumerate}
    \item Uses fewer resource-intensive gates (e.g., CNOT or $T$ gates) on average than the original circuit, and  
    \item Satisfies Eq. \eqref{eq:wee_cond}.  
\end{enumerate}  
While explicit algorithms exist for guaranteeing Eq. \eqref{eq:wee_cond} \cite{PhysRevA.95.042306}, they are difficult to scale to multi-qubit circuits and do not provide resource efficiency guarantees.

In the main text we have shown that both criteria can be met for several of the benchmarks we studied.   
Here, we present statistics that explain how this is achieved through quadratic reduction in the weighted ensemble error and selective filtering of circuit blocks.

\begin{figure*}[t]
    \centering
    \begin{subfigure}{0.7\textwidth}
        \centering
        \includegraphics[width=\linewidth]{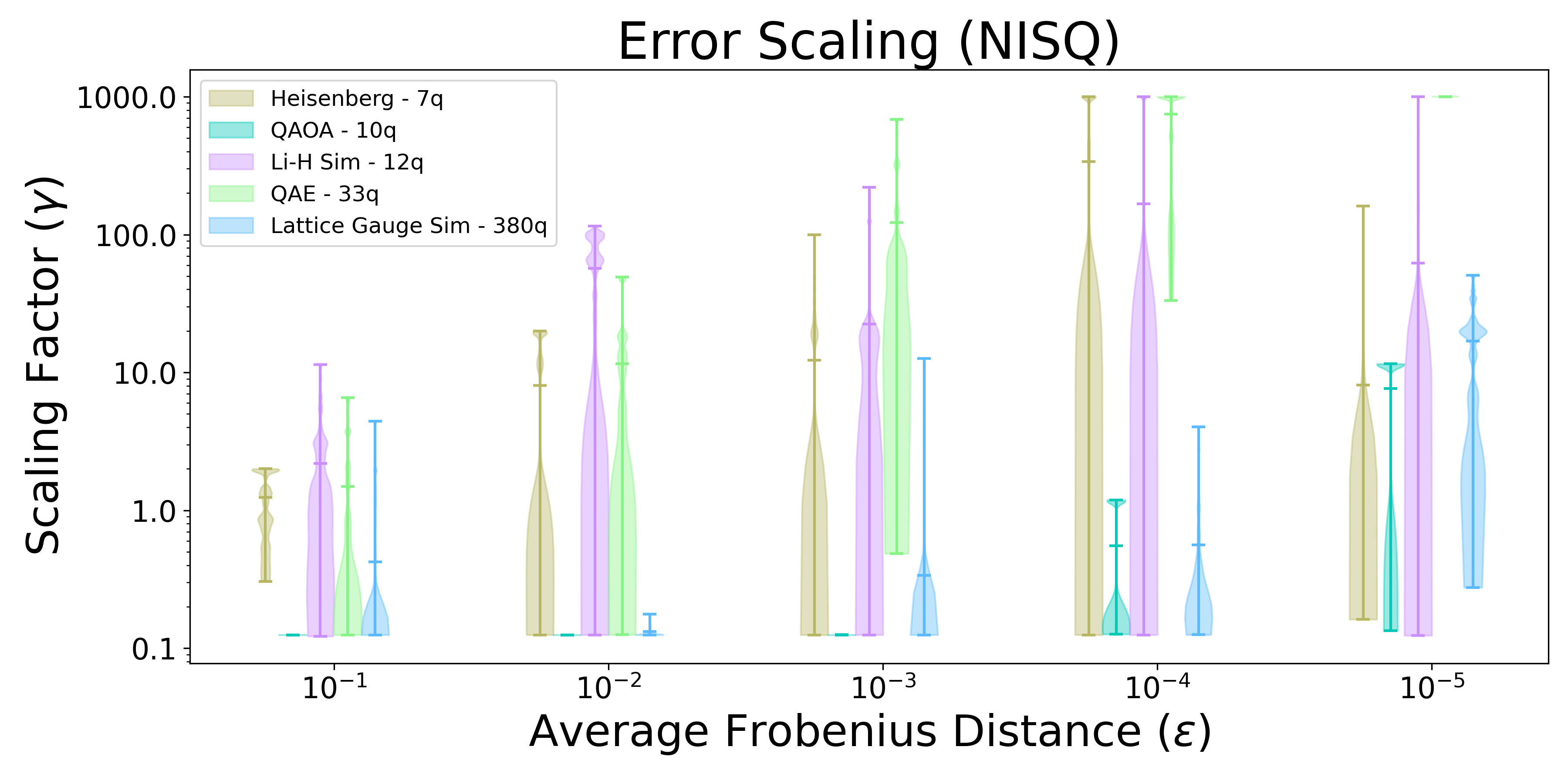}
    \end{subfigure} \\
    \begin{subfigure}{0.7\textwidth}
        \centering
        \includegraphics[width=\linewidth]{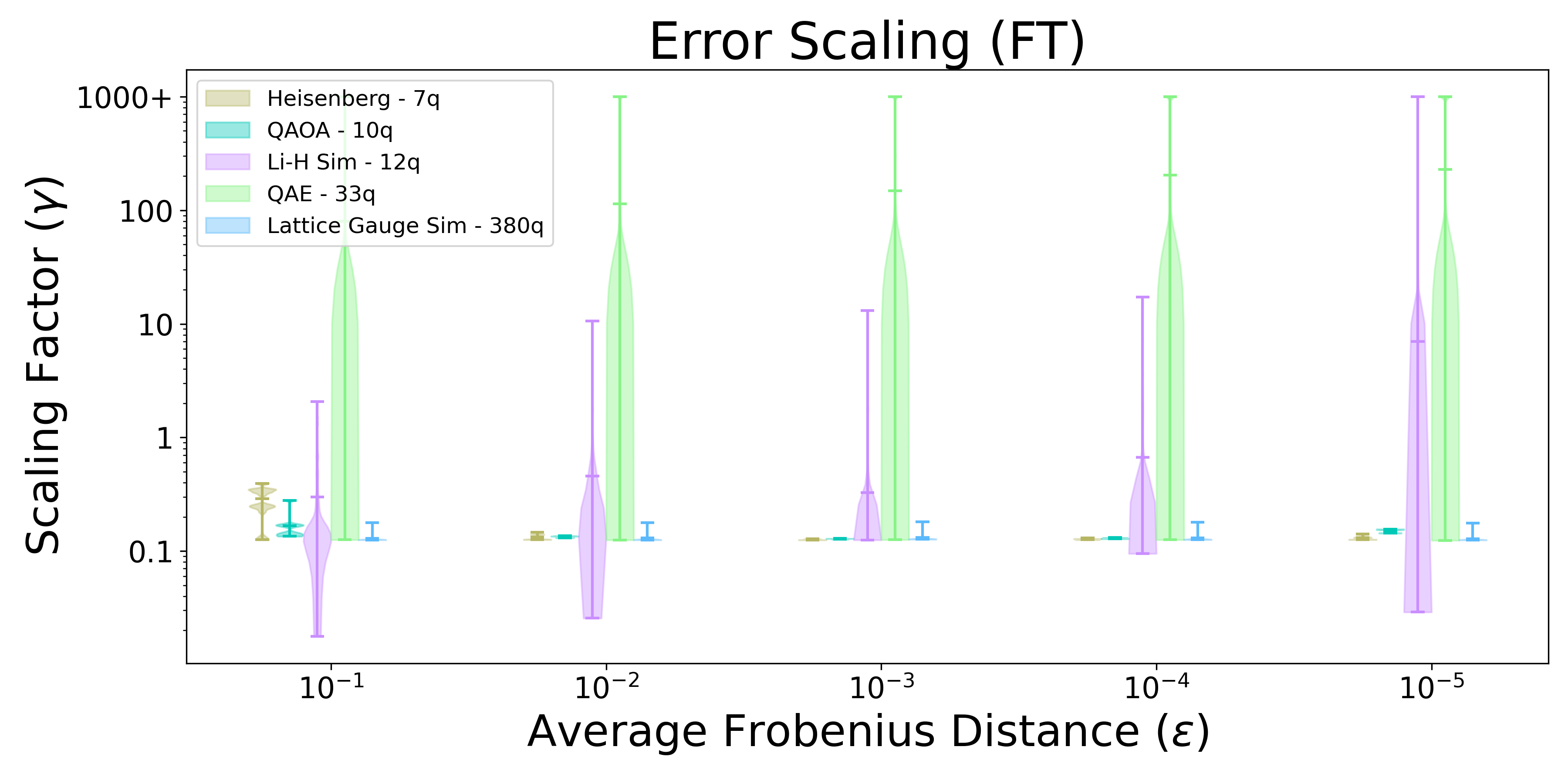}
    \end{subfigure}
    \caption{Quadratic scaling factor $\gamma$ as a function of $\epsilon$, for some of the benchmarks.  
    Top: NISQ workflow (minimizing CNOT count).  
    Bottom: FT workflow (minimizing $T$-count).}
    \label{fig:error_scaling}
\end{figure*}

\begin{figure*}[t]
    \centering
    \begin{subfigure}{0.7\textwidth}
        \centering
        \includegraphics[width=\linewidth]{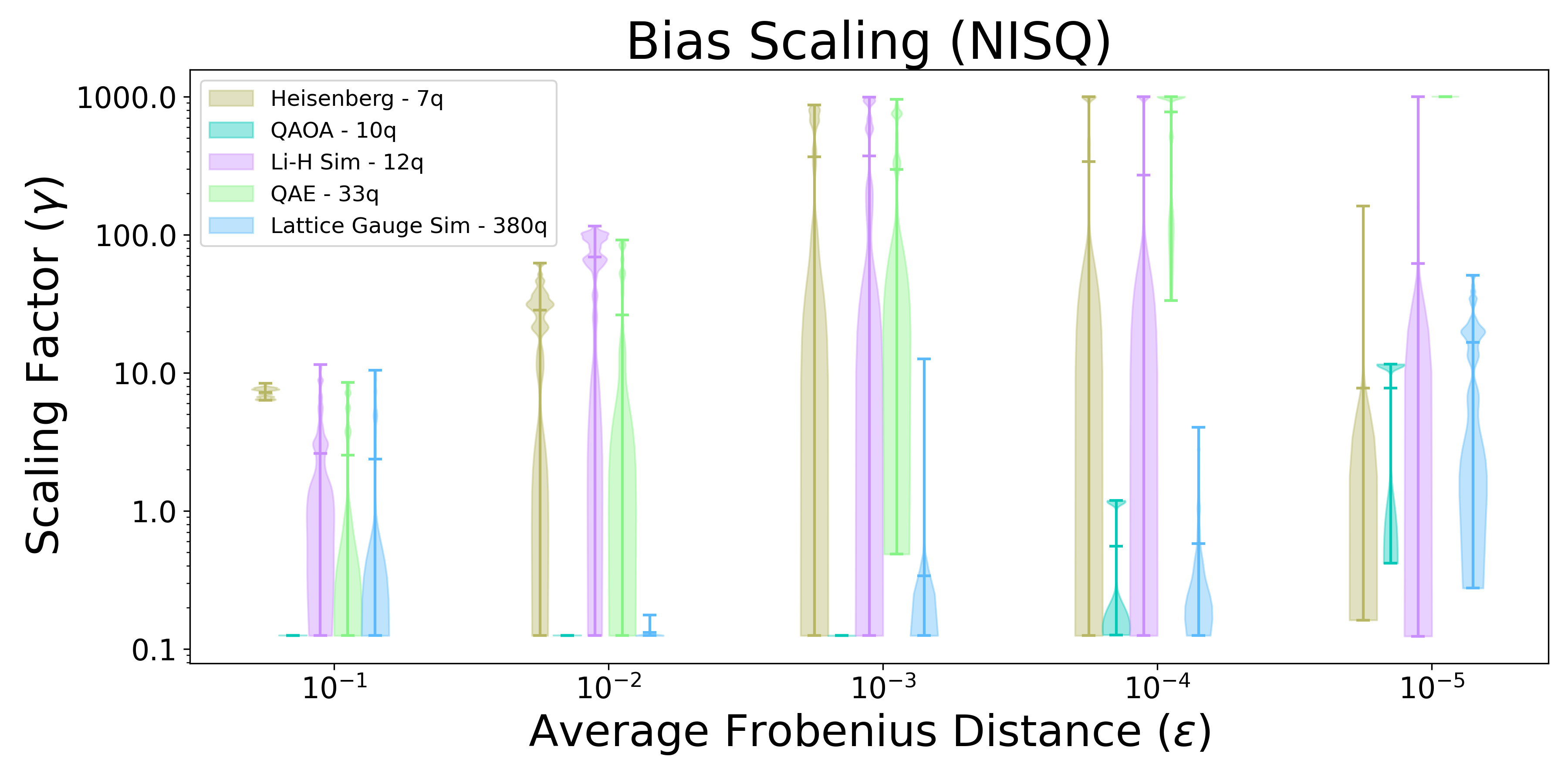}
    \end{subfigure} \\
    \begin{subfigure}{0.7\textwidth}
        \centering
        \includegraphics[width=\linewidth]{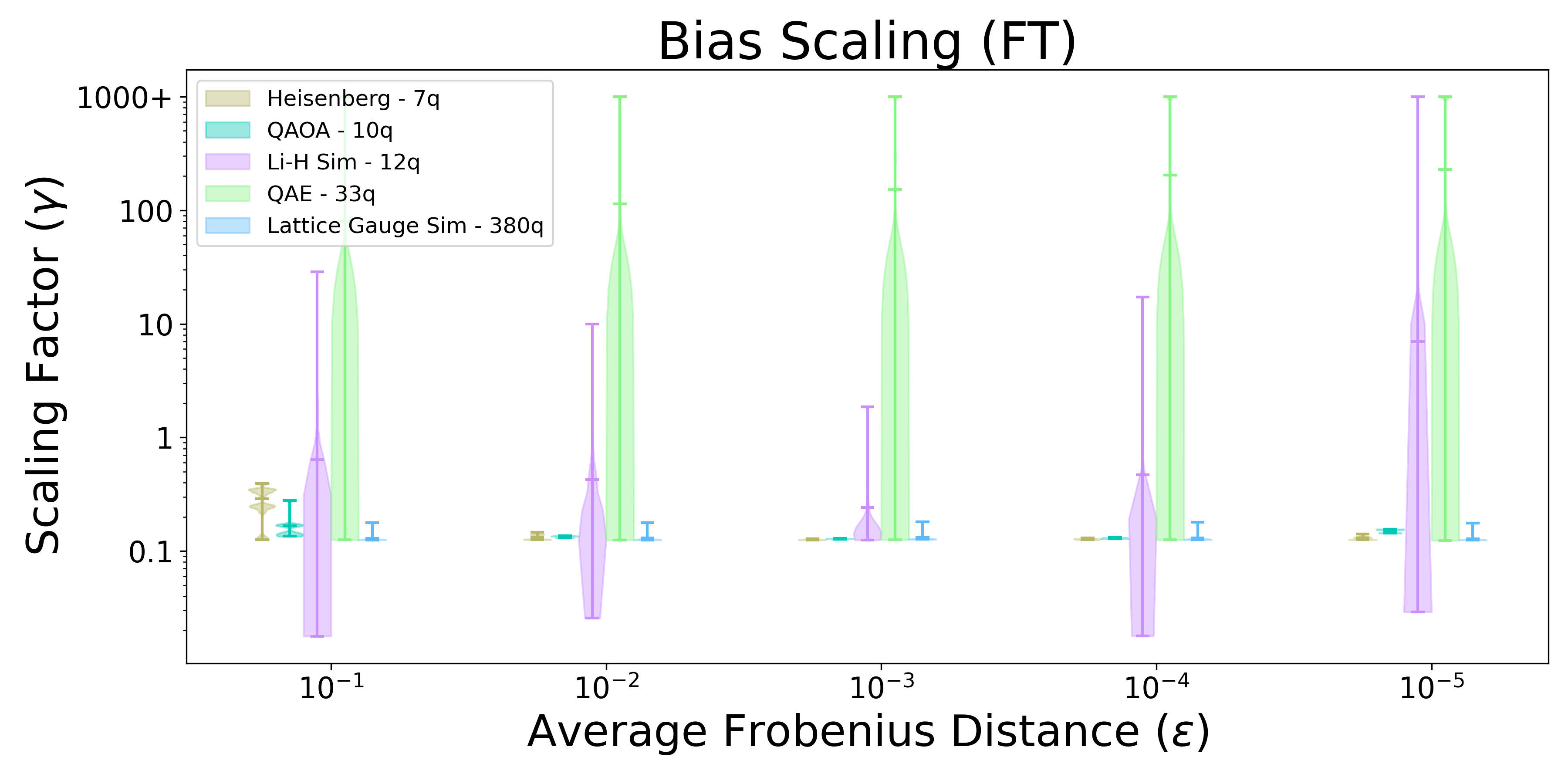}
    \end{subfigure}
    \caption{Quadratic scaling factor for the bias $\gamma_B$ as a function of $\epsilon$, for some of the benchmarks.  
    Top: NISQ workflow (minimizing CNOT count).  
    Bottom: FT workflow (minimizing $T$-count).}
    \label{fig:bias_scaling}
\end{figure*}

Figure \ref{fig:error_scaling} shows violin plots of the \emph{quadratic scaling factor} (QCF),  
\[
\gamma \equiv \frac{\normf{\sum_{i=1}^{M^\k} p^\k_i U^\k_i - V^\k}}{\epsilon^2},
\]  
for all blocks across many of our benchmarks.  
We expect $\gamma = O(1)$ when quadratic reduction of error is achieved.  
The plots indicate that, for most $\epsilon$ values, a significant subset of blocks in both NISQ and FT workflows achieve $O(\epsilon^2)$ scaling. In particular, almost all of the blocks in the FT workflow achieve $O(\epsilon^2)$ scaling. This is because we include a secondary compilation workflow, which trades off numerical synthesis resource reduction for improved scaling.

Our workflow does not require all blocks to satisfy Eq. \eqref{eq:wee_cond}. Blocks that fail to meet the condition are left unmodified, preserving the original circuit for those segments.  
This selective replacement explains why some benchmarks exhibit no gate count reduction (Tables 1 and 2 in the main text, particularly in the NISQ setting at small $\epsilon$). 

We also derived a sufficient condition for satisfying the ReWEE condition in terms of the bias of the compiler output,
\begin{align}
    \normf{\E{U} - V} < O(\epsilon^2).
\end{align}
We can define a quadratic scaling factor for the bias also, $\gamma_B \equiv \normf{\E{U}-V}/\epsilon^2$, and this quantity is displayed in Figure \ref{fig:bias_scaling} for all blocks of many of our benchmarks.

\section{Theoretical proofs}
In this section we provide proofs of Lemma 2 and Lemma 3 in the Methods section of the main text. Before restating the Lemmas and providing proofs, we recall several definitions from the main text.

$V$ is a target unitary on $n$ qubits, and $U_i$ are a set of approximate compilations of $V$ provided by a compiler. The compilation workflow can have probabilistic elements, and therefore we think of $U_i$ as independent, identically distributed (\emph{i.i.d.}) samples from some distribution of approximate compilations. The statistical properties we will demand of the compilation workflow are:
\begin{enumerate}
\item The compilations should all be close to the ideal, in the sense,
    \begin{align}
	\normf{U_i-V} \leq \epsilon, \quad \forall i		
		\label{eq:ass1}
	\end{align}
	\item The compilations have bounded variance, in the sense, $\E{U_i} \equiv \E{U}$ is independent of $i$, and
	\begin{align}
		\E{\normf{U_i - \E{U}}^2} \leq \epsilon', \quad \forall i
		\label{eq:ass2}
	\end{align}
\end{enumerate}
The expectations in these expressions and in the rest of this note are over the distribution of compilations generated by the compiler that satisfy Condition 1, \ie
$ \E{f(U)} = \int {\rm d}\mu_V(U) f(U), $
where ${\rm d}\mu_V(U)$ is the measure over $SU(2^n)$ defined by the compiler output that satisfies Condition 1 when the target is $V$ (the expectation symbol $\E{}$ should have a subscript $V$ but we omit this for ease). In addition,
\begin{enumerate}
    \item Define the channels $\mathcal{V}[\rho] = V \rho V\dg$, $\mathcal{U}_i[\rho] = U_i \rho U_i\dg$, and $\mathcal{U}[\rho] = \sum_i p_i U_i \rho U_i\dg$, for some probabilities $\{p_i\}$.
    \item $J_{\mathcal{E}} \equiv \sum_{i,j=1}^d \mathcal{E}(\ket{i}\bra{j})\otimes \ket{i}\bra{j}$ is the Choi matrix representation of the channel $\mathcal{E}$, which we assume to be acting acting on $d\times d$ density matrices and outputting density matrices of the same size.
\end{enumerate}

\noindent \emph{\bf Lemma 2:} \emph{Let $V$ and $U_i$ be defined as above. Given probabilities $\{p_i\}_{i=1}^M$, ($\sum_{i=1}^M p_i=1$), a sufficient condition for achieving $\E{\normf{\sum_{i=1}^M p_i U_i - V}} \leq O(\epsilon^2)$ is the \emph{reduced bias condition}: $\normf{\E{U}-V} \leq O(\epsilon^2)$.}

\noindent \textit{Proof:} First, note that $\E{\normf{\sum_{i=1}^M p_i U_i - V}} \leq O(\epsilon^2)$ follows directly from $\E{\normf{\sum_{i=1}^M p_i U_i - V}^2} \leq O(\epsilon^4)$ through an application of Jensen's inequality, and so we prove the latter statement.
\begin{align}
    \E{\normf{\sum_{i=1}^M p_i U_i - V}^2} & \leq \E{\normf{\frac{1}{M}\sum_{i=1}^M U_i - V}^2} 
	   = \normf{\overline{bias}}^2 + \frac{1}{M}\overline{var} + \left(1-\frac{1}{M}\right)\overline{covar}, 
       \label{eq:b-v-c} 
\end{align}
where,
\begin{align}
	\overline{bias} &= (\E{U} - V) \label{eq:bias} \\
	\overline{var} &= \frac{1}{M}\sum_{i=1}^{M} \E{\normf{U_i - \E{U}}^2} \label{eq:var} \\
	\overline{covar} &= \frac{1}{M(M-1)}\sum_{i=1}^M \sum_{j \neq i}^M \mathbb{E} \left\{ \tr[(U_i-\E{U})\dg \cdot (U_j - \E{U})] \right\} 
    \label{eq:covar}
\end{align}
The first inequality follows from the fact that the weights $p_i$ are optimized to reduce the ensemble error. The equality on the second line follows from using the definition of the Frobenius norm to decompose the quantity of interest in a \emph{bias-variance-covariance} decomposition, a well-known decomposition of generalization error of ensembles of estimators in machine learning \cite{Ueda_1996, Brown_20005}. Here, $\overline{bias}$ is the bias of the (uniform) ensemble, $\overline{var}$ is the (sample) average of the variance of the members of the ensemble, and $\overline{covar}$ is the covariance of the ensemble members. Since the ensemble members are sampled \emph{i.i.d.} from the compiler output distribution, the covariance term is zero. In addition, using the fact that the samples have bounded variance, Eq. \eqref{eq:ass2}, we get
$\E{\normf{\sum_{i=1}^M p_i U_i - V}^2} \leq \normf{\E{U}-V}^2 + \frac{\epsilon'}{M}$.
This quantity can be bounded as $O(\epsilon^4)$ by satisfying the reduced bias condition and choosing $M \geq \epsilon'/\epsilon^4$. \QED
\\

\noindent \emph{\bf Lemma 3:} \emph{Let $V$ and $U_i$ be defined as above, and let these unitaries act on a register of $n$ qubits; \ie ${\rm dim}~ V = {\rm dim}~ U_i = 2^n \equiv d$. Define $\mathcal{V}[\rho] = V \rho V^{\dagger}$ as a target channel, and $\mathcal{U}[\rho] = \sum_{i=1}^M p_i U_i \rho U_i\dg$ as an optimized ensemble channel, such that $d_\diamond(\mathcal{V}, \mathcal{U})\leq O(\epsilon^2)$. Further, let $\hat{\mathcal{U}}[\rho] = \frac{1}{T}\sum_{i=1}^T U_{\sigma(i)} \rho U_{\sigma(i)}\dg$ be the empirical channel defined by sampling and averaging over $U_i$ according to the distribution $\{p_i\}$ -- \ie $\sigma(i) \in \{1,...,M\}$. For $\epsilon, \delta \in (0,1)$, $\pr\left\{ d_\diamond(\mathcal{V}, \hat{\mathcal{U}}_T) \leq O(\epsilon^2) \right\} > 1-\delta$, given
\begin{align}
    T \geq \frac{2v + \frac{2}{3}R \left(\epsilon^2/(2d^2)\right) }{\left(\epsilon^2/(2d^2)\right)^2}\log \frac{2d^2}{\delta}
    \label{eq:T_bound}
\end{align}
Here, $v \equiv \normo{\sum_{i=1}^M p_i (J_i - J_{\mathcal{U}})^2}, R \equiv \max_i \normo{J_i - J_{\mathcal{U}}}$, and $J_i (and J_{\mathcal{U}}$) are the $d^2$-dimensional Choi matrices associated to the channel $\mathcal{U}_i[\rho] = U_i \rho U_i^{\dagger}$ (and $\mathcal{U}$).}

\noindent \textit{Proof:} We begin by an application of the triangle inequality:
\begin{align}
    d_\diamond(\mathcal{V}, \hat{\mathcal{U}}) &\leq  d_\diamond(\mathcal{V}, \mathcal{U}) +  d_\diamond(\mathcal{U}, \hat{\mathcal{U}}) \nn \\  
    & = O(\epsilon^2) + d_\diamond(\mathcal{U}, \hat{\mathcal{U}}),
    \label{eq:lemma3_1}
\end{align}
and hence we are left with the task of determining the sample size $T$ necessary to satisfy $d_\diamond(\mathcal{U}, \hat{\mathcal{U}}) \leq \epsilon^2$. Therefore, 
\begin{align}
    d_\diamond(\mathcal{U}, \hat{\mathcal{U}}) &\equiv \normd{\mathcal{U}- \hat{\mathcal{U}}} \nn \\
    & \leq \normt{J_\mathcal{U} - J_{\hat{\mathcal{U}}}} \nn \\
    & \leq 2d^2 \normo{J_\mathcal{U} - J_{\hat{\mathcal{U}}}},
\end{align}
where the first inequality follows from an identity established by Watrous \cite[Exercise 3.6]{watrous_2018}, the second is the application of a standard norm inequality ($\normt{A}\leq {\rm rank}(A)\normo{A}$), and an overestimate of the relevant rank: ${\rm rank}(J_\mathcal{U} - J_{\hat{\mathcal{U}}}) \leq {\rm rank}(J_\mathcal{U} ) + {\rm rank}(J_{\hat{\mathcal{U}}}) \leq 2d^2$. In the above, the Choi matrices are explicitly, $J_\mathcal{U} = \sum_{i=1}^M p_i J_i$ and $J_{\hat{\mathcal{U}}} = \frac{1}{T} \sum_{i=1}^T J_i$, with $J_i \equiv J_{\mathcal{U}_i}$. 

Now that the error is expressed in terms of an operator norm, we can apply standard Bernstein matrix concentration inequality \cite{tropp_2015} to get, 
\begin{align}
    \pr\left\{ d_\diamond(\mathcal{U}, \hat{\mathcal{U}}) \geq \epsilon^2 \right\} \leq \pr\left\{ \normo{J_\mathcal{U} - J_{\hat{\mathcal{U}}}} \geq \frac{\epsilon^2}{2d^2} \right\} \leq 2d^2 \exp\left(-\frac{T(\frac{\epsilon^2}{2d^2})^2}{2v + \frac{2}{3}R \frac{\epsilon^2}{2d^2}}\right),
\end{align} 
where $v$ and $R$ are the variance and maximum deviation as defined in the Lemma statement. In order for the right-hand side to be at most $\delta$, the sample size given in Eq. \eqref{eq:T_bound} is sufficient. Finally, combining this with the statement in Eq. \eqref{eq:lemma3_1} yields the Lemma statement. 
\QED

\subsection{Statistics of factors entering sample complexity}
While the formal sample complexity bound in Lemma 3 scales with $\epsilon$ as $O(1/\epsilon^4)$, in the main text we claimed that in practice, $v=O(\epsilon^2)$, and therefore, the number of samples needed is $O(1/\epsilon^2)$. Figure 5 in the main text provided evidence for this gentler scaling for one of the benchmarks in our study. While Lemma 3 is stated in terms of diamond distance between the empirical channel and the ensemble channel, this error is infeasible to evaluate numerically because of the difficulty of calculating the diamond distance for channels acting on many qubits. Therefore, in Figure 5 of the main text, we showed the trace distance error in the output, $\normt{\rho_{T} - \rho_i}$,
where $\rho = \sum_{i=1}^M p_i U_i \rho_{0} U_i\dg$ is the output of the ensemble channel for random input state $\rho_{0}$, and $\rho_{T} = \frac{1}{T}\sum_{i=1}^T U_i \rho_{0} U_i\dg$ is the output of the empirical channel for the same input (the $U_i$ in the empirical channel are sampled according to the distribution $\{p_i\}$ set by the ensemble channel). 

\begin{figure*}[ht]
    \centering
        \includegraphics[width=0.7\linewidth]{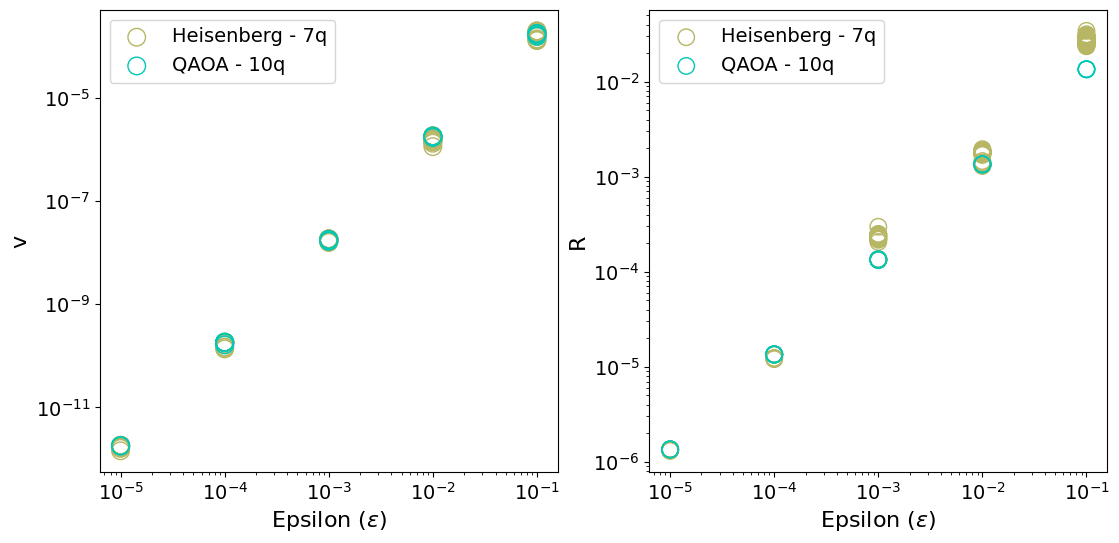}
    \caption{Values of variance and maximum deviation factors, $v$ and $R$, entering the sample complexity bound Eq. \eqref{eq:T_bound_state} for each block of two benchmark circuits.}
    \label{fig:vR}
\end{figure*}

The equivalent sample complexity bound for this error metric is $\pr\{\normt{\rho_{T} - \rho} \geq \epsilon^2\} \leq 1-\delta$, if
\begin{align}
    T \geq \frac{2v + \frac{2}{3}R \left(\epsilon^2/(2d)\right) }{\left(\epsilon^2/(2d)\right)^2}\log \frac{2d}{\delta},
    \label{eq:T_bound_state} 
\end{align}
where $v \equiv \normo{\sum_{i=1}^M p_i (\rho_i - \rho)^2}, R \equiv \max_i \normo{\rho_i - \rho}$, and $\rho_i = U_i\rho_0 U_i\dg$. Calculating the values $v$ and $R$ is difficult for most of the benchmarks we studied, even in this case, because $M$ can be very large for full circuits, and thus computing the ideal ensemble output $\rho$ is numerically intensive. Instead, we compute these variance and maximum deviation values, $v$ and $R$, for all blocks in two of our benchmark circuit ensembles. As shown in Fig. \ref{fig:vR}, $v \ll \epsilon^2$ in practice, which explains the gentler scaling of $T$ required to approximate the ensemble channel with the empirical channel.

